\documentclass[iop]{emulateapj}

\usepackage{graphicx} 
\usepackage{subfigure} 

\slugcomment{ApJ, accepted}

\shorttitle{Near-IR IFU Observations of SMGs}
\shortauthors{Olivares et al.}

\begin{document}

\title{Spatially Resolved Spectroscopy of Sub-mm Galaxies at $z$$\simeq$2\footnote{Based on observations made with ESO Telescopes at the La Silla Paranal Observatory under programme IDs 088.A-0452 and 090.A-0464}}

\author{V. Olivares\altaffilmark{1,11}, E. Treister\altaffilmark{1,2,11}, G. C. Privon\altaffilmark{1,2,11}, S. Alaghband-Zadeh\altaffilmark{3},  Caitlin M. Casey\altaffilmark{4,5}, 
K. Schawinski\altaffilmark{6},  P. Kurczynski\altaffilmark{7}, E. Gawiser\altaffilmark{7}, N. Nagar\altaffilmark{1,11}, S. Chapman\altaffilmark{3},  F. E. Bauer\altaffilmark{2,8,9,11}, 
D. Sanders\altaffilmark{10}}
\altaffiltext{1}{Universidad de Concepci\'{o}n, Departamento de Astronom\'{\i}a, Casilla 160-C, Concepci\'{o}n, Chile}
\altaffiltext{2}{Instituto de Astrof\'{\i}sica, Facultad de F\'{i}sica, Pontificia Universidad Cat\'{o}lica de Chile, Casilla 306, Santiago 22, Chile}
\altaffiltext{3}{Institute of Astronomy, University of Cambridge, Madingley Road, Cambridge, CB3 0HA UK}
\altaffiltext{4}{Department of Physics and Astronomy, University of California, Irvine, CA 92697, United States}
\altaffiltext{5}{Department of Astronomy, The University of Texas at Austin, Austin, TX 78712, United States}
\altaffiltext{6}{Institute for Astronomy, Department of Physics, ETH Zurich, Wolfgang-Pauli-Strasse 27, CH-8093 Zurich, Switzerland}
\altaffiltext{7}{Department of Physics and Astronomy, Rutgers, The State University of New Jersey, 136 Frelinghuysen Road, Piscataway, NJ 08854, USA}
\altaffiltext{8}{Millennium Institute of Astrophysics, MAS, Nuncio Monse\~{n}or S\'{o}tero Sanz 100, Providencia, Santiago de Chile} 
\altaffiltext{9}{Space Science Institute, 4750 Walnut Street, Suite 205, Boulder, Colorado 80301} 
\altaffiltext{10}{Institute for Astronomy, 2680 Woodlawn Drive, University of Hawaii, Honolulu, HI 96822, USA}
\altaffiltext{11}{EMBIGGEN Anillo}

\begin{abstract}
We present near-infrared integral-field spectroscopic observations targeting H$\alpha$ in eight  sub-millimeter galaxies (SMGs) 
at $z$=1.3-2.5 using VLT/SINFONI, obtaining significant detections for six of them. The star formation rates derived from the H$\alpha$ emission are $\sim$100 M$_\odot$yr$^{-1}$, 
which account for only $\sim$ 20-30\% of the infrared-derived values, thus suggesting that these systems are very dusty. Two of these systems present [NII]/H$\alpha$ ratios indicative of the presence of an Active Galactic 
Nucleus (AGN). We mapped the spatial distribution and kinematics of the star forming regions in these galaxies on kpc-scales. In general, the H$\alpha$ morphologies tend to be highly irregular 
and/or clumpy, showing spatial extents of $\sim$3-11~kpc. We find evidence for significant spatial offsets, of $\sim$0.1-0.4$''$ or 1.2-3.4 kpc, between the H$\alpha$ and the continuum emission 
in three of the sources. Performing a kinemetry analysis we conclude that the majority of the sample is not consistent with disk-like rotation-dominated kinematics. Instead, they tend to show 
irregular and/or clumpy and turbulent velocity and velocity dispersion fields. This can be interpreted as evidence for scenario in which these extreme star formation episodes are triggered by 
galaxy-galaxy interactions and major mergers. In contrast to recent results for SMGs, these sources appear to follow the same relations between gas and star forming rate densities as less luminous and/or normal star forming galaxies.
\end{abstract}


\section{Introduction}

Early observations of the sub-mm sky, mostly carried out using the Submillimeter Common-User Bolometer Array (SCUBA) on the James Clerk 
Maxwell Telescope (JCMT), revealed a population of very luminous, L$_{IR}$$>$10$^{12}$L$_\odot$ --- corresponding to associated star-formation rates (SFR) 
$\geq$500M$_\odot$yr$^{-1}$ ---, high redshift, $z$$>$1, galaxies which are extremely faint at optical wavelengths (e.g., \citealp{smail97,hughes98}), the so-called 
sub-mm galaxies (SMGs). These sources, as a population, are responsible for the release of a significant fraction of the energy generated by all galaxies over the history 
of the Universe \citep{blain99}. Their very high IR to sub-mm fluxes, in comparison to the optical and UV outputs, indicate that a significant fraction of the luminosity in these 
galaxies is heavily obscured \citep{casey14b}. Intriguingly, SMGs contribute about 20\% \citep{chapman05} to the integrated star formation rate 
density ($\rho$SFR), thus indicating that a large fraction of the star formation in the Universe is hidden by gas and dust \citep[e.g.,][]{barger12}.

Mostly due to their optical faintness, measuring redshifts and assigning accurate multi-wavelength associations have been difficult and expensive in terms of
telescope time. There is strong evidence that the majority of the SMGs are at redshifts greater than unity, with a median redshift for the 
population of $z$$\simeq$2-3 \citep{smail00,smail02, chapman05, simpson14}. Accurate spectroscopic redshifts and well-determined multi wavelength
properties have now being derived for a few hundred SMGs \citep*[see][for a review]{casey14}. The observed gas and stellar surface densities of these SMGs, when 
coupled to plausible spectral energy distributions (SEDs), are significantly higher than those of low-redshift IRAS galaxies \citep{blain99,trentham99,dunne00}.
Furthermore, their luminosities and masses demand that the SMG phase be short-lived as compared with the age of the Universe \citep[e.g.,][]{frayer99}.
The typical depletion timescale in SMGs measured from molecular gas CO observations is $\sim$100-200 Myr, in contrast to the much longer, $\sim$1 Gyr, 
timescales derived for normal galaxies \citep[e.g,][]{tacconi10, engel10, bothwell13, swinbank14}. Still, the timescales for high redshift SMGs are longer than those 
seen in local ULIRGs, primarily due to the elevated gas fractions in high-redshift galaxies \citep{greve05, bothwell13}.

SMGs are mostly found at $z$$>$1, and Luminous ($L_{IR}$$>$10$^{11}$$L_\odot$) and Ultra-luminous ($L_{IR}$$>$10$^{12}$$L_\odot$) 
IR Galaxies, (U)LIRGs \citep{sanders96} are considered their analogues in the local Universe. However, while (U)LIRGs in the local Universe are extremely rare, with 
space densities $\sim$10$^{-7}$Mpc$^{-3}$ (comparable to those of quasars; \citealp{kim98}), SMGs at $z$$\simeq$2.5 have space densities $\sim$10$\times$ higher at similar 
IR luminosities \citep[e.g.,][]{chapman05, simpson14}. This clearly points to a rapid evolution in the most extreme IR luminous galaxies, which has been parameterized as (1+$z$)$^\alpha$ 
with $\alpha$$\simeq$3.5 for the total energy output attributed to these systems out to z$\sim$1 \citep{lefloch05}. Hence, while they are relatively unimportant in the local Universe, (U)LIRGs 
can account for up to $\sim$50\% of the total star formation density by $z$$\sim$2 \citep[e.g.,][]{murphy11,casey12,gruppioni13}.

The nature of the large energy output in SMGs is now commonly assumed to be associated with their vigorous star formation activity. Nuclear activity is also expected to provide a contribution
to the energy emission in a fraction of the systems. Early studies \citep[e.g.,][]{alexander05b} claimed that a very high fraction, $\sim$75\% of the radio-selected spectroscopically-identified 
SMGs, can be associated with Active Galactic Nuclei (AGN). In contrast, a lower AGN fraction, $\sim$15-30\%, was reported for the general SMG population \citep{laird10,johnson13,wang13}, the 
vast majority of them subject to heavy obscuration \citep{bauer10}.  A strong connection between AGN activity and strong starburst episodes has been proposed based both on observations 
\citep{sanders88} and theory \citep{hopkins08}, in which these events are triggered by the major merger of gas rich galaxies. Spatially-resolved observations of SMGs using the 
\textit{Hubble Space Telescope} have not been conclusive in this regard \citep{swinbank10,chen15}, finding that SMGs at $z$=0.7-3.4 are in general compact with spheroid/elliptical light profiles. 
Similarly, recent ALMA observations \citep[e.g.,][]{simpson15} have also found that brighter SMGs are compact, $\sim$0.2$''$, in the sub-mm as well. 

Integral Field Unit (IFU) spectrographs, in particular those assisted by Adaptive Optics (AO), have been critical in studying the morphologies and kinematics of SMGs at high spatial resolutions. 
The first comparison of the dynamical properties from SINFONI (Spectrograph for Integral Field Observations in the Near Infrared) installed at the VLT (Very Large Telescope) by \citet{bouche07} 
showed that bright SMGs have larger velocity widths and are much more compact than the UV and optically selected $z$$\sim$2 star forming galaxies; they also have lower angular momenta and 
higher matter densities. This indicates that dissipative major mergers may dominate the SMG population. Early studies in the near-IR by \citet{swinbank06} of 6 high redshift SMGs show that they 
consist of more than one sub-component with significant velocity offsets. Similar results were found by \citet{alaghband-zadeh12} for eight SMGs, who interpreted them as evidence that SMGs are 
systems in the early stages of major mergers. More recently, \citet{menendez-delmestre13} found evidence for the presence of a compact region characterized by broad H$\alpha$ emission which 
can be associated with an AGN together with clumps of narrow H$\alpha$ emission due to star formation processes. At the same time, they do not see evidence for ordered motions, hence 
consistent with a connection between SMGs and galaxy mergers. \citet{harrison12} observed eight ULIRGs that host AGN activity and SMGs at high redshift with IFU, concluding that 
the AGN activity could be the dominant power source for driving all of the observed outflows, although star formation may also play a significant role in some of the sources. Unfortunately, given 
the relatively low spatial densities of SMGs and the required exposure times using 8m telescopes and state-of-the-art instrumentation, only relatively small samples of galaxies have been studied 
so far.

In this paper, we present observations of eight SMGs between $z$=1.4 and 2.5 using the SINFONI near-IR IFU at the European Southern Observatory (ESO) Very Large Telescope (VLT). 
The main goal of this work is to study the morphology and kinematics of the star forming regions in high redshift SMGs using the H$\alpha$ line
as a tracer. In Section \ref{sec:obs} we present the properties of the sample studied, the observations, data reduction and analysis. The results of these 
observations are presented in Section \ref{sec:results}, while the discussion and conclusions are presented in Section \ref{sec:discussion} and \ref{sec:conclusions} respectively. 
We assume a $\Lambda$CDM cosmology  with $h_0$=0.7, $\Omega_m$=0.27 and $\Omega_\Lambda$=0.73 \citep{hinshaw09}.

\section{VLT/SINFONI Near-IR IFU Observations}
\label{sec:obs}

\subsection{Sample}

As described in the previous section, only relatively small samples of targeted SMGs at high-z have been studied so far using IFU
spectroscopy. This paper focuses on follow-up IFU observations for eight $z$$>$1 ULIRGs selected based on two different and complementary criteria.
Hence, it will be possible to study how these selections are related to the properties of the observed sources.
The basic properties of the sources studied in this work are presented in Table~\ref{tab_properties}. We categorize our sources in two groups, Group A focuses on 
warm-dust ULIRGs, while in Group B we include only SMGs with a suitable reference star for the AO corrections to have high spatial resolution for 
kinematic measurements.

\begin{deluxetable*}{ccccccccccccc}
\tablecaption{Sample Properties.\label{tab_properties}}
\tablecolumns{12}
\tablehead{ 
\colhead{Name} &
\colhead{Group\tablenotemark{a}} &
\colhead{RA } &
\colhead{Dec} &
\colhead{$z$$_{spec}$} &
\colhead{Date} &
\colhead{Band} &
\colhead{Seeing} &
\colhead{FOV} &
\colhead{t$_{exp}$} &
\colhead{L$_{FIR}$\tablenotemark{b}} &
\colhead{T$_{dust}$.\tablenotemark{c}} \\
\colhead{} &
\colhead{} &
\multicolumn{2}{c}{(J2000)} &
\colhead{} &
\colhead{} &
\colhead{} &
\colhead{[Arcsec]} &
\colhead{} &
\colhead{[Hrs]} &
\colhead{[10$^{13}$ L$_{\odot}$]} &
\colhead{[K]} 
}

\startdata
\noalign{\smallskip}
J033129.874-275722.40 & A &03:31:29.87 & -27:57:22.40 & 1.482 & 11/2011 & H & 0.80 &8$''$$\times$8$''$& 2.0  &1.2$_{-0.4}^{+0.7}$& 45$\pm$4 \\
J033212.866-274640.89 & A &03:32:12.86 & -27:46:40.89 & 1.930  & 11/2011& K & 0.64 &8$''$$\times$8$''$ & 2.5  & 2.2$_{-1.4}^{+4.3}$& 42$\pm$9\\
J033246.329-275327.01  & A &03:32:46.32 & -27:53:27.01 & 1.382  &10/2011 & H & 0.65 &8$''$$\times$8$''$ & 2.0  &  0.4$_{-0.1}^{+0.1}$& 53$\pm$13\\
J033249.352-275845.07  &  A& 03:32:49.35 & -27:58:45.07 & 2.326  &10/2011& K & 0.86 &8$''$$\times$8$''$ &  2.0 & 8.1$_{-2.6}^{+3.9}$& 56$\pm$5\\
L50879 & A & 03:33:37.67 & -27:46:35.26 & 2.509  & 11/2011& K & 0.64 &8$''$$\times$8$''$ & 2.5  & 0.8 & 46  \\
RGJ030258+001016  & B &  03:02:58.94 & +00:10:16.3 & 2.236  & 10/2012 & K &  0.34 &3$''$$\times$3$''$ &   3.75  & 0.77$_{-0.14}^{+0.14}$& $>$46 \\
SMMJ04431+0210 (N4)  & B	& 04:43:07.25 &  +02:10:23.3 & 2.509  & 10/2012 & K & 0.35 &3$''$$\times$3$''$  & 3.75  & 0.3 & 40 \\
SMMJ2135-0102  & B & 21:35:11.60 & -01:02:52.0 & 2.326  & 10/2012 & K & 0.34 &3$''$$\times$3$''$  & 3.75 & 0.23$_{-0.02}^{+0.02}$ &30-60 \\
\enddata
\tablenotetext{a}{Group A correspond at no-AO observation of warm-dust ULIRGs, while Group B are AO observations focus on SMGs.}

\tablenotetext{b}{FIR luminosity (8-1000 $\mu$m) of J033129, J033212, J033246 and J033249 from \citet{casey11a}, for RGJ0302+0010 from \citet{swinbank04}, for SMMJ04331+0210 (N4) from \citet{neri03} --- corrected for 
lensing magnification and obtained from 850$\mu$m flux densities --- and for SMMJ2135-0120 from \citet{ivison10}, also corrected for lensing magnification.}

\tablenotetext{c}{T$_{dust}$ were obtained from \citet{casey11a} for J033129, J033212, J033246 and J03329, for RGJ0302+0010 from \citet{chapman04}, for SMMJ04331+0210 (N4) from \citet{neri03} and for 
SMMJ2135-0120 from \citet{ivison10} with two components associated with the warm and cool dust.}
\end{deluxetable*}

\subsubsection{Group A -- $250~\mu m$ Selected}

Group A consists of several warm-dust star-forming ULIRGs at $z$$\sim$1-2. 
While SMGs are proposed to contribute to as much as half of the star formation density at early epochs \citep{gruppioni13}, SMG observations at 850~$\mu$m 
naturally select colder-dust ULIRGs with T$_{d}$$<$45K and thus potentially miss a whole subpopulation of high-z ULIRGs by virtue of their warmer dust temperatures. Recent 
work by \citet{casey09,casey11a} show that there is a population of ULIRGs that are fainter at 850~$\mu$m and have higher dust temperatures, T$_{d}$$\simeq$50 K \citep[e.g.,][]{chapman04, blain04}. These 
so-called warm-dust ULIRGs may contribute significantly to the cosmic star formation rate density at its peak, similar to the SMG contribution. However, the warmer-dust 
temperatures implied by the absence of 850 $\mu$m-flux suggests that they might have inherently different evolutionary origins compared to 
the well-studied cold-dust SMGs. It is possible that these higher temperatures are due to the effects of AGN heating.

We have assembled a sample of warm-dust star-forming ULIRGs from spectroscopic samples of 250~$\mu$m-selected BLAST galaxies from the observations 
described by \citet{casey11a}. All of them have existing redshift measurements from long-slit spectroscopy. In four of the five galaxies (J033246, J033249, J033129 
and J033212) the redshift was determined based on the detection of the H$\alpha$ emission line, while for L50879 the Ly$\alpha$ line was detected and identified.

J033246 was first identified in spectroscopic surveys by \citet{kriek08} and \citet{vanzella08}. It was later classified as a star-forming radio galaxy based
on a significant sub-mm detection by \citet{weiss09}. The galaxy has a redshift of $z$=1.382, as indicated by the detection of an emission line
using VLT/ISAAC identified as H$\alpha$ by \citet{casey11a}. The measured rest-frame [NII]/H$\alpha$ line flux ratio of 0.13 suggests that
the emission in this galaxy is dominated by star formation activity. J033249 is a very similar source, also classified as a vigorous star forming galaxy
at $z$=2.326 \citep{casey11a}. Two systems (J033212 and J033129) in our sample were not detected by our VLT/SINFONI observations and thus
are not further discussed.

\subsubsection{Group B -- SMGs with natural AO Observations}

Sources in this group were selected roughly independently of the target physical properties. The main goal in this case is to achieve the highest possible spatial resolution for our kinematic measurements. Hence, the 
only SMG selection criteria were the availability of a suitable reference star for the AO corrections (which for VLT/SINFONI was a R$<$14 star closer than 30$''$), a measured spectroscopic redshift of 
$z$$\sim$1.5 or $z$$\sim$2.5, so that H$\alpha$ would fall in either the H or K near-IR bands, and that the source is observable at high enough elevation by the VLT.

This resulted in the selection of three sources for this group: SMMJ2135-0102, RGJ0302+0010 and SMMJ04431+0210 (N4). The mean redshift for these sources is $z$=2.189.
SMMJ04431+0210 (N4) is one of the 15 original sub-mm sources found in the SCUBA Lens Survey (S$_{850}$=7.2 mJy; \citealp{smail97, smail02}). SMMJ2135-0102 was 
identified by \citet{swinbank10} in a APEX/LABOCA 870 $\mu$m observation of the galaxy cluster MACSJ2135-010217. The galaxy RGJ0302+0010 
was selected from the catalogs of SMGs of \citet{chapman04,chapman05,takata06}. 

\paragraph{RGJ030258+001016}
was identified as a ULIRG by \citet{chapman04}. Rest-frame UV spectroscopy shows strong [CIV] emission, which together with 
a high [NII]/H$\alpha$ suggests the presence of AGN activity and a redshift of 2.2404$\pm$0.0008 \citep{chapman04,swinbank04}. The  
[O III]$\lambda\lambda$4959,5007 emission-line doublet shows a narrow and redshifted broad component. This broad component is offset from the center by about 8 kpc and can 
be explained by an outflow with its near-side obscured by dust \citep{harrison12}. 

\paragraph{SMMJ04431+0210 (N4)}
is an extremely red galaxy identified as the near-IR counterpart of the faint sub-mm source reported by \citealt{smail99}, which was 
undetected in deep optical images \citep{smail98}. It is located behind the $z$=0.18 cluster MS0440+02, with an amplification factor for the background galaxy of $\mu$=4.4 \citep{smail99}.
A spectroscopic redshift of 2.5092$\pm$0.0008 was measured by \citet{frayer03} based on the detection of an H$\alpha$, [NII]$\lambda$$\lambda$6583,6548, and 
[OIII]$\lambda$5007 emission lines using the NIRSPEC spectrograph at Keck. The rest-frame [NII]/H$\alpha$ line flux ratio of 0.47$\pm$0.06 
suggests that this emission can be explained by a narrow-line AGN/LINER nucleus surrounded by a resolved starburst.

\paragraph{SMMJ2135-0102}
(also known as the ``cosmic eyelash'') is a gas-rich starburst galaxy. The galaxy was identified in the field of the massive lensing cluster MACSJ2135-010217 ($z$$_{cl}$=0.325) via 
ground-based 870$-$$\mu$m imaging \citep{swinbank10}. They unambiguously identified the redshift of the source as $z$=2.3259$\pm$0.0001 thanks to the detection of 
carbon monoxide (CO) J=1$-$0 emission at 34.64 GHz and derive an amplification factor for the background galaxy of $\mu$=32.5$\pm$4.5. 
Later, \citet{swinbank11} used the IRAM Plateau de Bure Interferometer and the EVLA to obtain maps of the CO(6-5) and CO(1-0) emission, 
finding that the molecular gas kinematics are well described by a rotationally-supported disk with an inclination-corrected rotation speed 
v$_{rot}$=320$\pm$25 km s$^{-1}$, a ratio of rotational to dispersion-support of v/$\Sigma$=3.5$\pm$0.2  and a dynamical mass of 
(6.0$\pm$0.5)$\times$10$^{10}$ M$_{\odot}$ within a radius of 2.5 kpc. Follow-up sub-mm studies of both molecular and atomic transitions in this
galaxy reported by \citet{danielson11} and \citet{danielson13} found evidence for a two-phase medium and physical properties such as ISM density and far-UV radiation fields 
similar to those observed in nearby ULIRGs and central regions of starburst galaxies.

\subsection{Observations and Data Reduction}

Near-IR IFU observations of the sample described above were performed using the SINFONI instrument \citep{Eisenhauer03}, installed at the ESO VLT. These data
were obtained as part of program 088.A-0452 in October and November 2011 for sample A and program 090.A-0464 observed from October 2011 to November 2012 for sample B.

For the observations of the group A sources we used SINFONI in no-AO mode, with a 250 milli-arcsec pixel scale and a field-of-view (FOV) 
of 8$''$$\times$8$''$. For group B, sources were selected to have a suitable AO reference star, R$<$14 mag closer than 30$''$. 
Hence, observations were taken using the natural guide star (NGS) AO mode, using the 100 milli-arcsec pixel scale and a 3$''$$\times$3$''$ FOV. 
Since the main goal of these observations was to map the H$\alpha$ emission, we used either the H-band grating, which provides 
a wavelength coverage of 1.45-1.85 $\mu$m and a spectral resolution of 3000, or the K-band grating with a spectral resolution of 4000 and
wavelength coverage of 1.95 to 2.45 $\mu$m, depending on the redshift of the source. The K grating was used for all sources except for J033246. The total integration time 
per source ranged from 2 to 4 hours. Since each source occupied a small fraction of the FOV, the surrounding empty regions were used for sky-subtraction,
thus achieving a 100\% on-source time. Each observation was broken into 1-hour observing blocks (OB), which includes all overheads. For each OB, a corresponding telluric
standard star was observed. 

The primary data reduction was performed using the ESO SINFONI pipeline version 2.5.2\footnote{\scalebox{0.7}{Available at http://www.eso.org/sci/software/pipelines/sinfoni/sinfoni-pipe-recipes.html}.} This 
pipeline is organized as a set of 6 stand-alone recipes. The first step evaluates the detector linearity; this is produced by analyzing flat fields taken with increasing intensity in order to create a 
map of highly non-linear bad pixels. Then, a master dark frame and a map of hot pixels are created from the median of a series of dark frames. Finally, flat field frames with nearly constant 
intensity are combined to produce a master flat field and a third bad-pixels map (BPM). These three BPMs are then combined to produce a master BPM. The pipeline then computes the geometric 
distortions by placing a fiber at different positions on the detector. Each science frame is therefore corrected for dark current, bad pixels, geometric distortions and then flat fielded and wavelength calibrated.

We then performed sky estimation and subtraction, which is critical when dealing with faint sources, such as those in our sample, at near-IR wavelengths. Since our targets have relatively 
small spatial extents, the SINFONI FOV contains enough source-free area to provide a good sky estimation. The pipeline provides two methods to estimate and subtract the sky contribution. The 
first one, which computes the median of all images on a set as an estimation of the sky, is not very accurate especially when the sky is not stable. The second method, which we chose for our data 
reduction, uses the closest frame in time as an estimation of the sky to be subtracted.  All frames are then shifted and co-added. They contain both spectral and spatial information, and are therefore 
reconstructed to produce 3D cubes. 

The extraction of a 2D image from the data cube for each source was done by creating a pseudo-longslit using the QfitsView software\footnote{\scalebox{0.7}{Available at http://www.mpe.mpg.de/$\sim$ott/QFitsView/}} and 
its ImRed analysis option. The size of the extraction aperture for each source is determined as a compromise between the maximum gathered flux and the minimum sky residual contribution. Then, the 
apall \textsc{IRAF} task was used to extract an integrated 1D spectrum for each source. Finally, additional sky correction, particularly useful to eliminate residuals of sky 
emission lines, was carried out using the Skycorr package \citep{noll14}. Flux calibration and telluric corrections were performed using observations of standard stars closer than 2 hours in time and
with a $\Delta$$z$$<$1.2 in airmass. Specifically, we used the Fitting Utility for SINFONI (FUS) package developed by Dr. Krispian Lowe as part of his PhD thesis\footnote{\scalebox{0.7}{Available at 
http://uhra.herts.ac.uk/bitstream/handle/2299/2449/Krispian\%20Lowe.pdf}}.

In order to estimate the spatial resolution of our observations, for each target we fitted Gaussian profiles to the images of the corresponding standard stars. As presented in Table 1, the effective angular 
resolution for the non-AO targets ranges from 0.64$''$ to 0.84$''$, with a mean and median of 0.7 and 0.65 respectively (corresponding to $\sim$0.59 kpc and $\sim$0.55 kpc at z$\sim$2). For the 
galaxies in group B, which correspond to AO-assisted observations, these values are: 0.34$''$ for mean and median (corresponding to $\sim$0.28 kpc at z$\sim$2). We note that these 
estimates come from shorter exposures ($\sim$10s) of brighter sources than the science exposures and are therefore most likely lower limits to the spatial resolution in our targets. However, 
we expect this underestimation to be rather small, $\sim$0.1$''$, and thus does not affect any of our conclusions.

\section{Results}
\label{sec:results}

With the SINFONI data described above, we were able to study both the galaxy-integrated and the spatially resolved properties of the H$\alpha$ emission 
in our galaxy sample. In this section, we present images of the near-IR continuum and H$\alpha$ emission, star formation rates, and test for evidence of AGN activity. We further 
study the star formation rate surface density, velocity fields and velocity dispersion maps. In Table~\ref{tab_spec_meas}, we present the spectroscopic properties for the sources in our sample derived 
from the VLT/SINFONI data, including H$\alpha$ fluxes, H$\alpha$ luminosities amongst others. The derived sources properties and associated SFRs, including those derived from IR observations 
obtained from the literature are listed in Table~\ref{tab_spec_der}. 

\subsection{Source Images}

\begin{figure*}
\centering
\subfigure{\includegraphics[width=.49\textwidth]{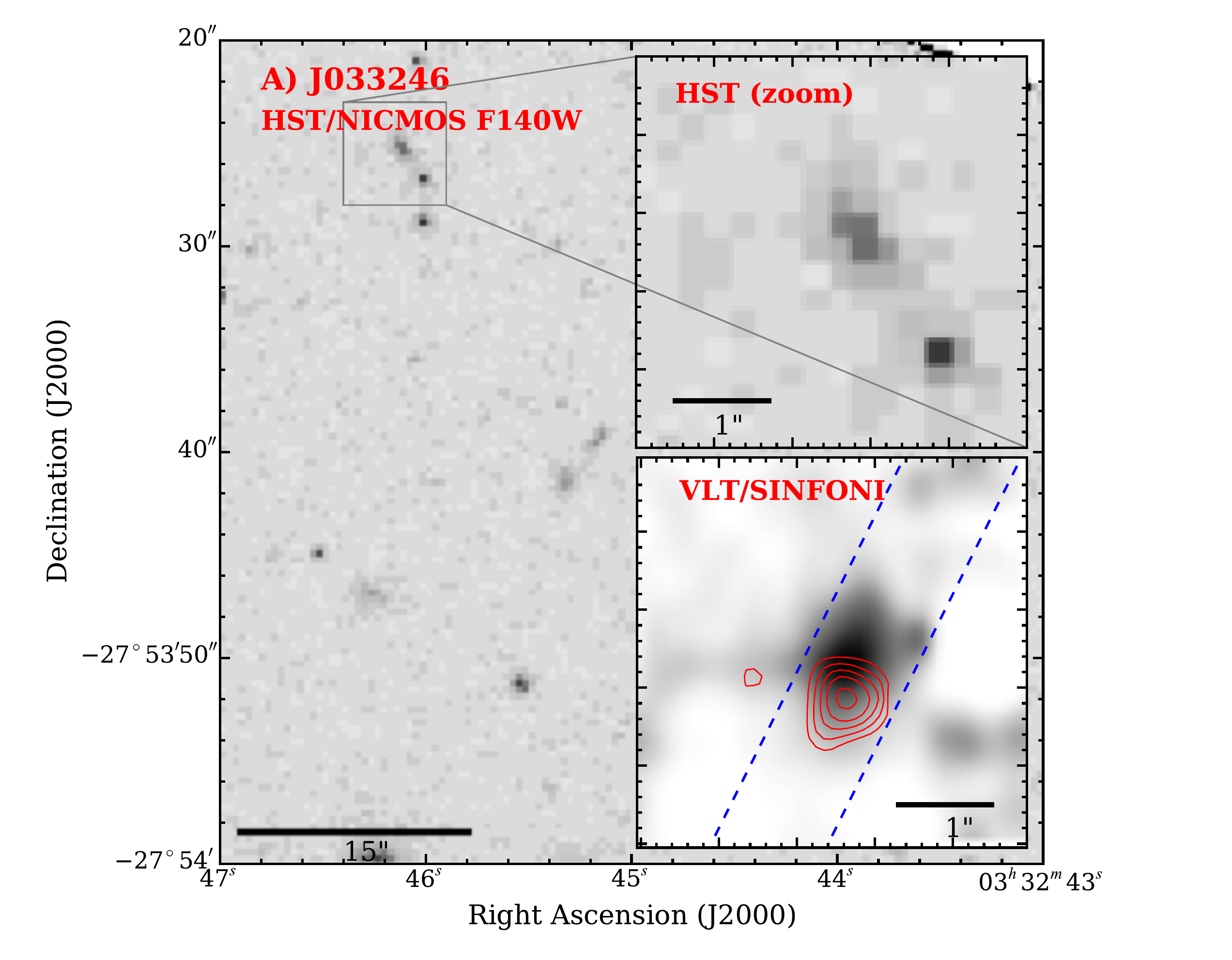}\label{J033246_img}}
\subfigure{\includegraphics[width=.49\textwidth]{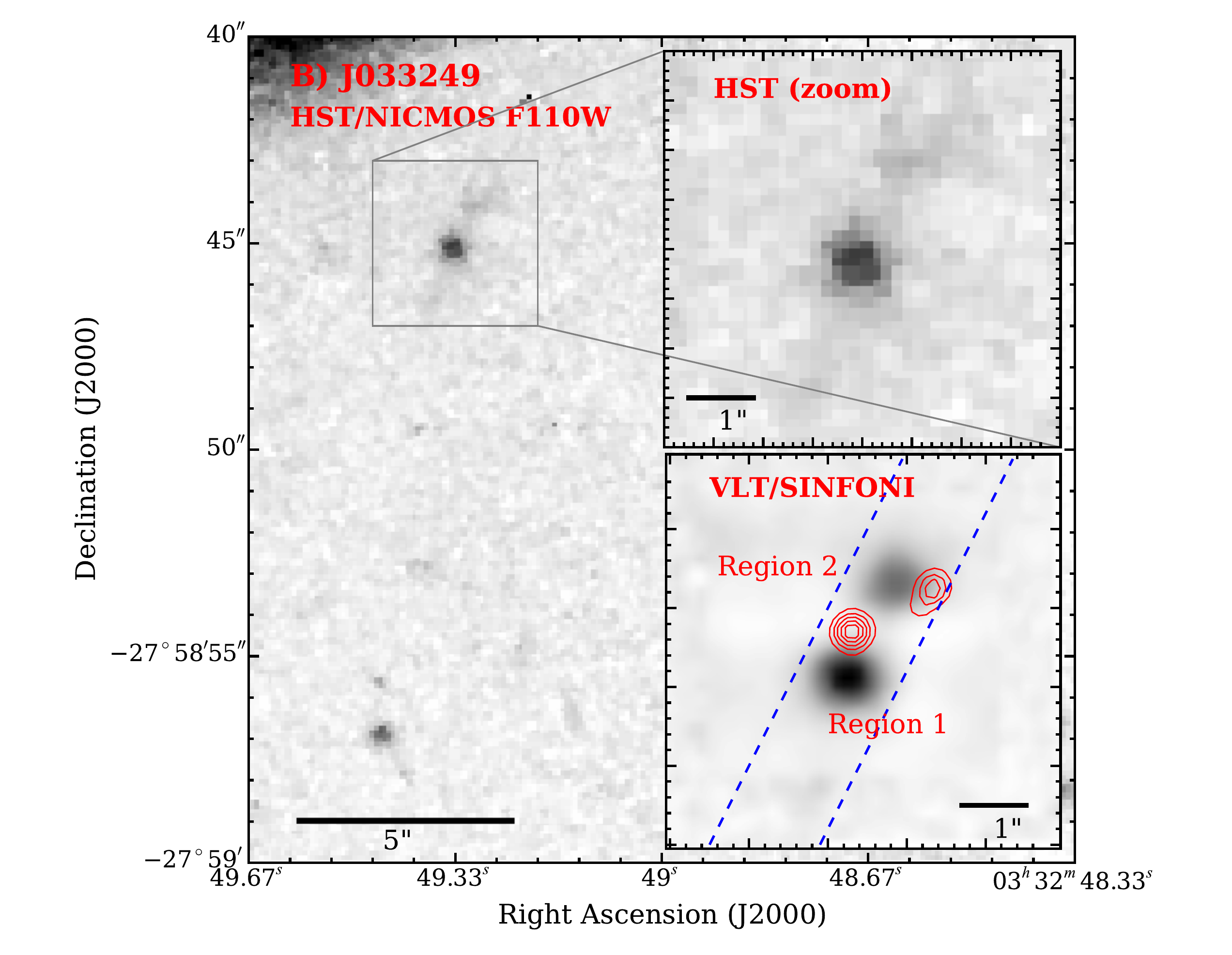}\label{J033249_img}}\\
\subfigure{\includegraphics[width=.49\textwidth]{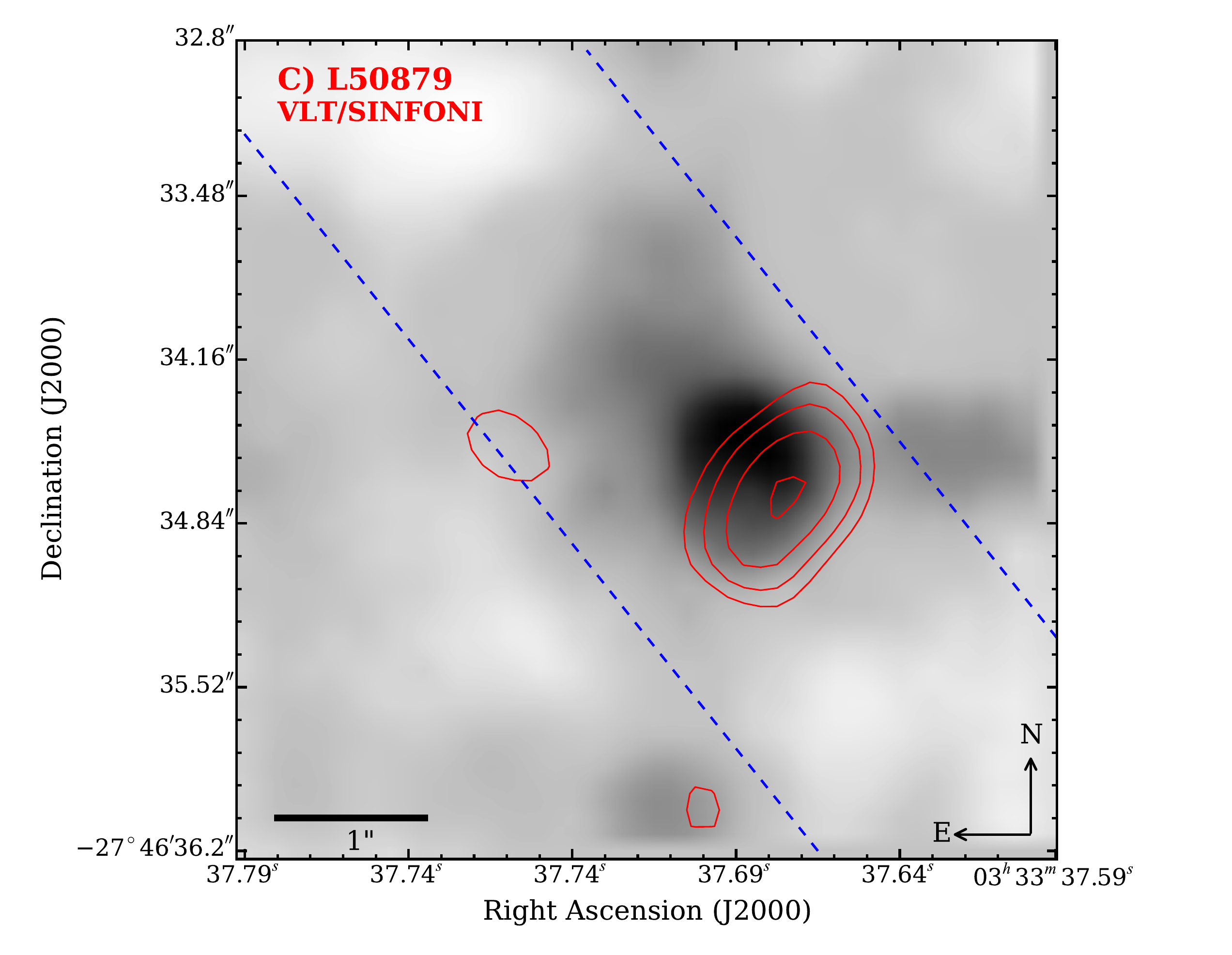}\label{L50879_img}}
\subfigure{\includegraphics[width=.49\textwidth]{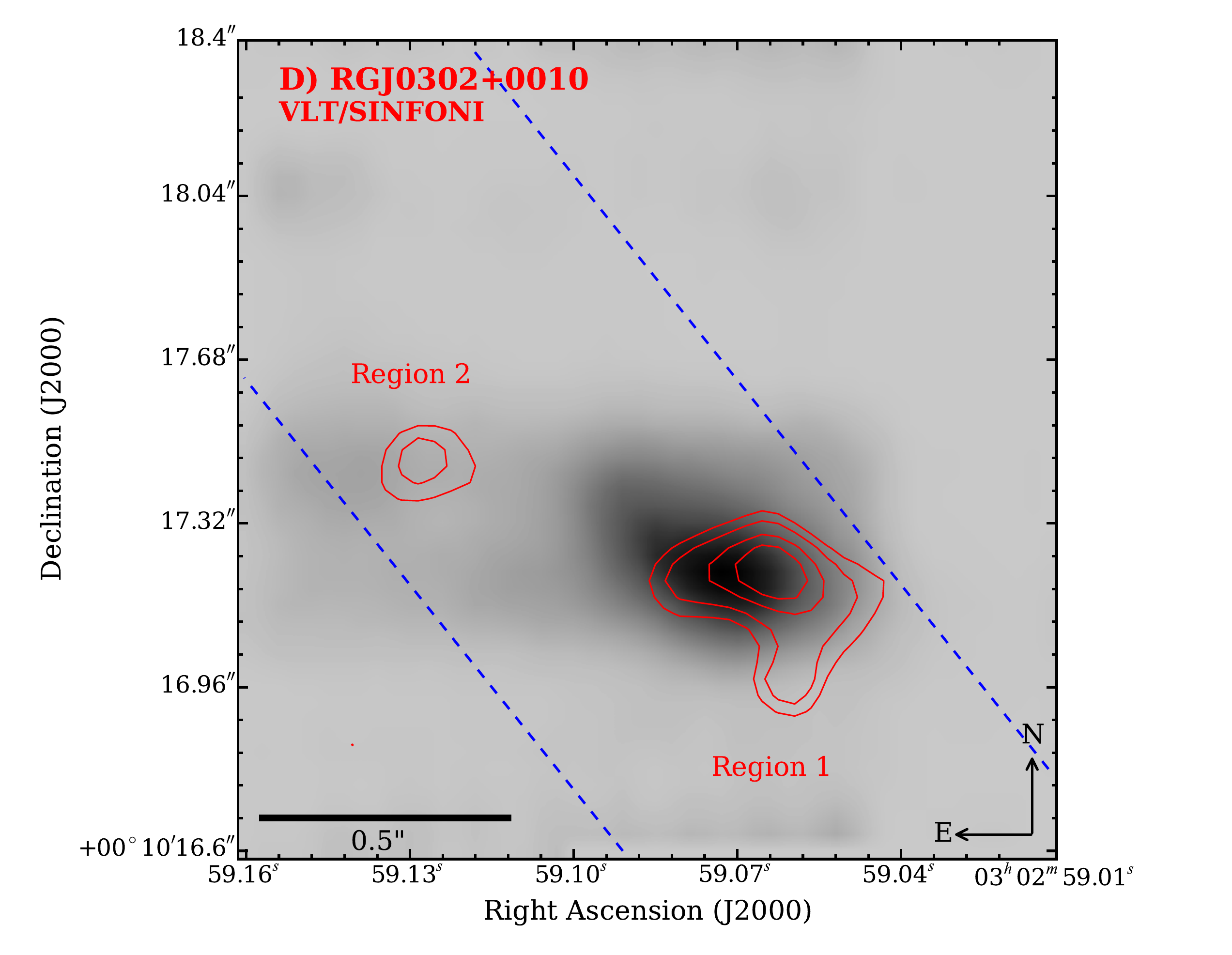}\label{RGJ0302_img}}\\
\subfigure{\includegraphics[width=.49\textwidth]{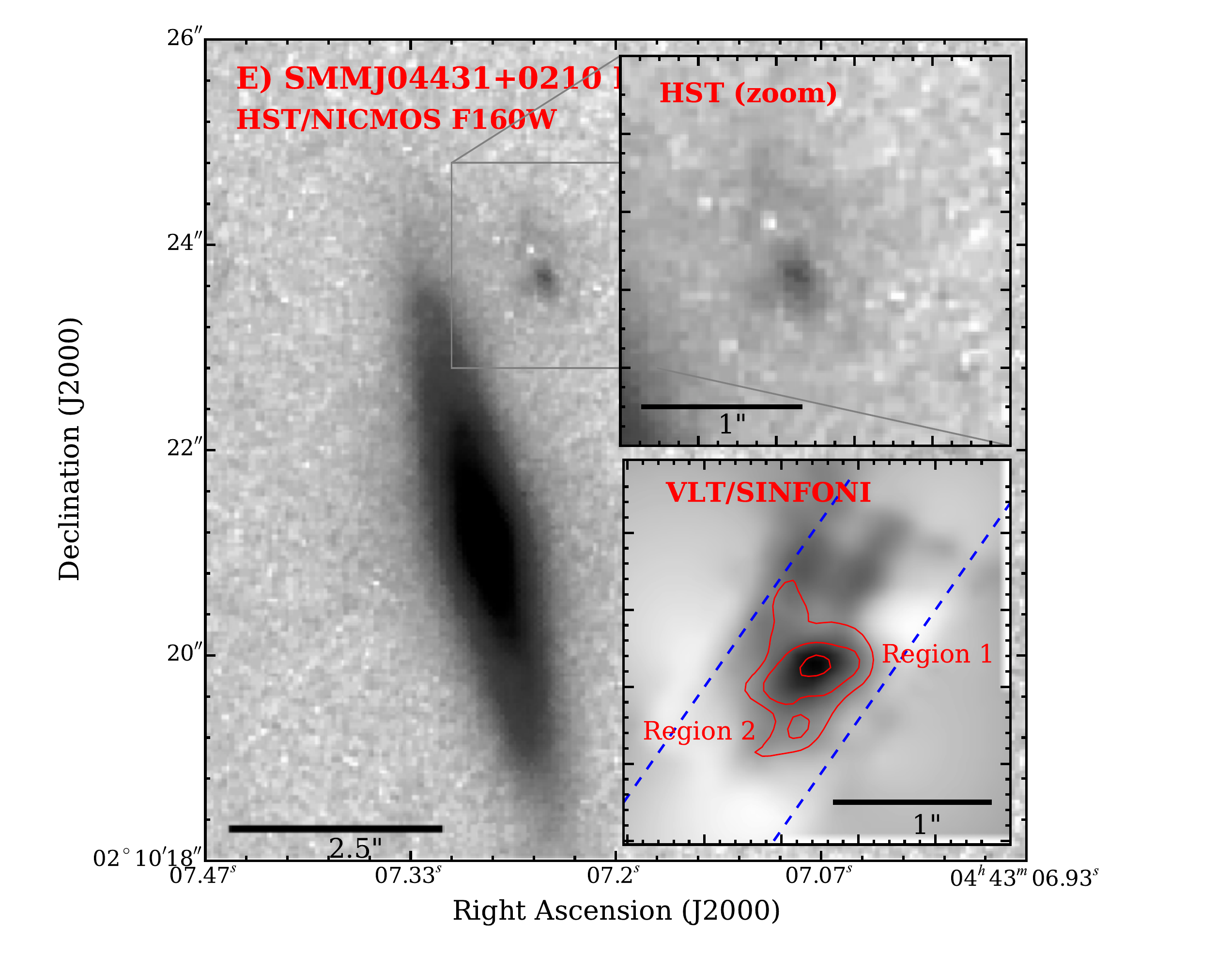}\label{smmj04431_img}}
\subfigure{\includegraphics[width=.49\textwidth]{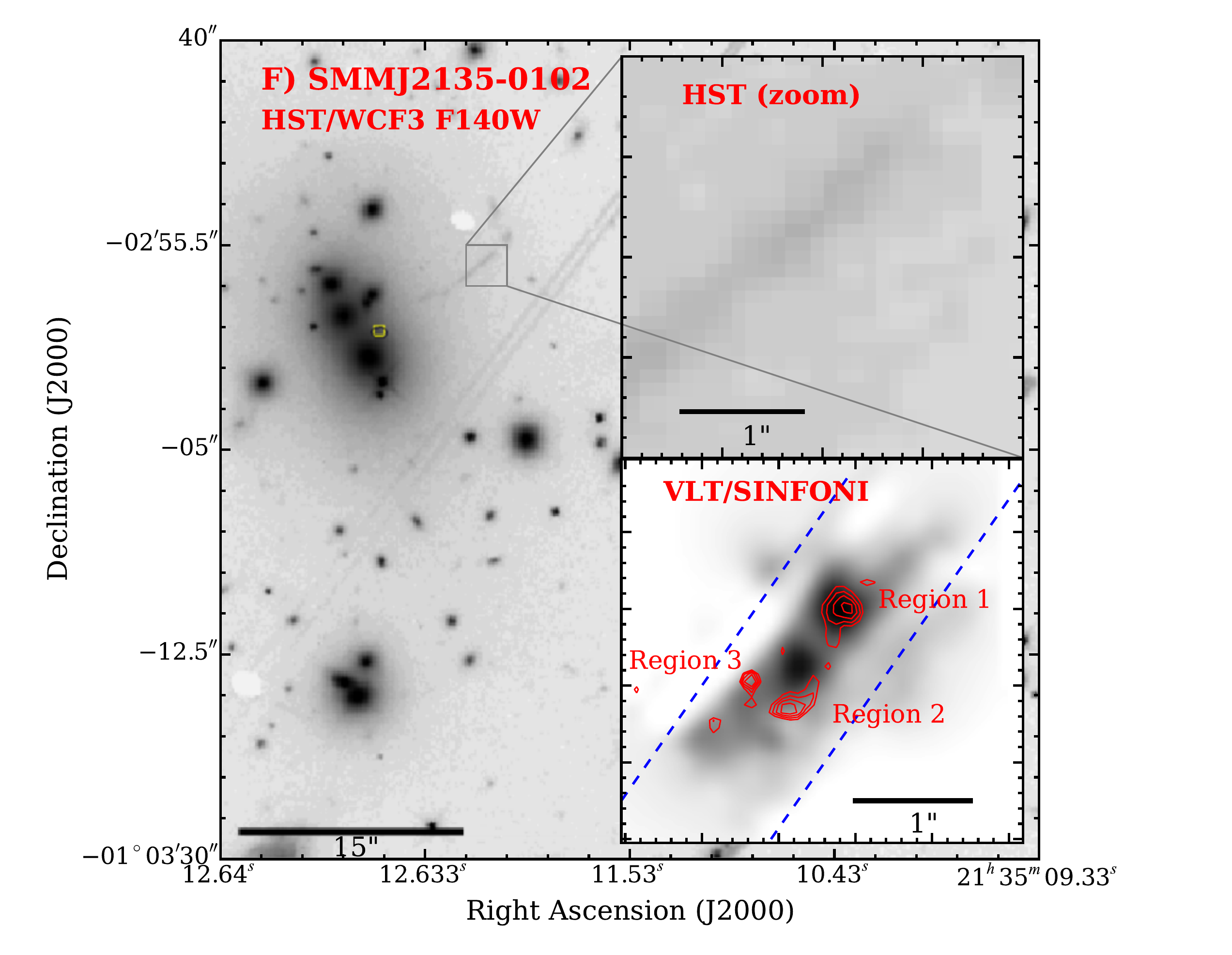}\label{smmj2135_img}}
\caption{Maps of the H$\alpha$ distribution for six sources in our sample: J033246, J033249, L50879, RGJ0302+0010, SMMJ04431+0210 (N4) and SMMJ2135-0102. The  
blue dashed lines show the position of the pseudo-longslits used in each of the targets. Where {\it HST} data are available, top insets correspond to zoom-ins around the region 
of the sub-mm emission, while the bottom inset show the reconstructed 2D images obtained from the VLT/SINFONI data cubes. In each case, red contours on the SINFONI images 
correspond to the H$\alpha$ emission. {\it J033246}: HST/NICMOS image in the F140W filter. H$\alpha$ inner/outer contours correspond to 6 to 2$\sigma$. 
{\it J033249}: HST/NICMOS F110W image. H$\alpha$ inner/outer contours correspond to 6 to 2$\sigma$. {\it L50879} image in the K band, contours correspond to 5 to 2$\sigma$.
{\it RGJ0302+0010}: SINFONI image in the K band, contours correspond to 4 to 2$\sigma$.  {\it SMMJ04431+0210}: HST/NICMOS image using the F160W filter. Contours correspond 
to 4 to 2$\sigma$. {\it SMMJ2135-0102}: HST/WCF3 F140W image of the massive cluster MACS J2135-0102, including the region of the SMMJ2135-0102 (the ``cosmic eyelash''). Contours 
corresponds to 5 to 2$\sigma$.}
\label{source_images}
\end{figure*}

Figure~ \ref{source_images} shows the \textit{HST} observed-frame near-IR images for J033246 and J033249, SMMJ04431+0210 (N4) and
SMMJ2135-0102. In all of them we also present a zoom-in around the region where the sub-mm emission is detected and a reconstructed 2-D image obtained from the 
VLT/SINFONI data cube. For RGJ0302 and L50879 only the VLT/SINFONI image is presented, as these sources were not observed by \textit{HST}. In all cases, the H$\alpha$ 
emission is overlaid on the VLT/SINFONI image as red contours.

\paragraph{J033246} 
J033246 presents an extended morphology with an angular extension of 1$\farcs$87, which corresponds to a physical size of 16 kpc. The H$\alpha$ emission is marginally resolved spatially,
with an angular diameter of 1$\farcs$69$\pm$0$\farcs$1, which corresponds to a physical size of 14.5$\pm$0.8 kpc. Interestingly, the region where the H$\alpha$ emission originates from appears to be significantly
displaced from the rest-frame optical continuum. This is further discussed in the next section. 

\paragraph{J033249} 
J033249 also shows an extended morphology, with an angular extension of $\sim$1$\farcs$47 and a size of 11.3 kpc for Region 1 and for $\sim$1$\farcs$28 or 10.7 kpc the Region 2, as can be seen in 
Fig~\ref{J033249_img}. The spatial distribution of H$\alpha$ emission is resolved and offset from the galaxy center, with an angular extension of $\sim$0$\farcs$82$\pm$0$\farcs$1 or 6.8$\pm$0.9 kpc for Region 1 and $\sim$0$\farcs$7$\pm$0$\farcs$1 
or 5.8$\pm$0.9 kpc for Region 2. The morphology of this system is consistent with that of an early-stage merger with a projected distance between the two components of 11.4 kpc.

\paragraph{L50879} 
L50879 presents a rather smooth morphology with a spatial extension of $\sim$2.32$''$, which corresponds to $\sim$10.25 kpc at the redshift of the source. The spatial distribution of 
the H$\alpha$ line emission shows that most of it is offset from the center of the system as discussed in the following section. For the H$\alpha$ emission we measured an angular size 
of $\sim$1$\farcs$2$\pm$0$\farcs$2  or $\sim$10$\pm$1.4 kpc.

\paragraph{RGJ0302+0010} 
RGJ0302+0010 shows a disturbed morphology, as can be seen in Fig~\ref{RGJ0302_img}. The brighter region has a spatial extension of $\sim$1$\farcs$15  or 9.8 kpc, 
while for the fainter one the size is $\sim$0$\farcs$35  or 3 kpc. The H$\alpha$ distribution is clumpy, presenting a larger central region (Region 1) of $\sim$0$\farcs$5$\pm$0.2 in size, corresponding to $\sim$4.2$\pm$2.0 kpc. Another 
region can be identified (Region 2), with an angular size of $\sim$0$\farcs$38$\pm$0.2 or 3.2$\pm$2.0 kpc.

\paragraph{SMMJ04431+0210 (N4)} 
For SMMJ04431+0210 (N4), the SINFONI/VLT K band image of this region reconstructed from the data cube shows a rather clumpy morphology, with one strong and prominent source and at least
one and possibly two other clumps. The H$\alpha$ emission is strongly concentrated on the southern, brightest, region. The H$\alpha$ emission is spatially resolved, showing a disturbed morphology. We can 
identify two marginally-separated components. The brighter one (Region 1) has a angular size of $\sim$0$\farcs$85 corresponding to 7.0 kpc, while the other 
one (Region 2) has a size of $\sim$0$\farcs$48 (3.9 kpc). The overall H$\alpha$ emission extends for $\sim$1$\farcs$4$\pm$0$\farcs$3, corresponding to 11.5$\pm$2.4 kpc, uncorrected for lensing amplification. 
Correcting for lensing magnification using a magnification factor of $\mu$=4.4 \citep{smail99}, Region 1 has a size of $\sim$1.6$\pm$0.3 kpc and Region 2 of $\sim$0.9$\pm$0.3 kpc.

\paragraph{SMMJ2135-0102} 

Figure~\ref{smmj2135_img} presents the HST/WCF3 F140W image of the region around the source SMMJ2135-0102 (the ``cosmic eyelash''), located near the massive cluster MACS J2135-0102. The 
K-band image obtained from the VLT/SINFONI data cubes shows a clumpy morphology, in which at least two bright star-forming regions in the source plane can be clearly seen, corresponding to regions identified as 
Y1 and Y2 by \citet{swinbank11}. These regions are separated by $\sim$2.8 kpc and have a spatial extension of $\sim$0$\farcs$5, corresponding to 2.4 kpc. We can further distinguish a smaller unresolved star forming region, 
labeled ``region 3''. For the region Y1 we measure an angular size of $\sim$0$\farcs$6$\pm$0$\farcs$1 corresponding to a diameter of $\sim$2.8$\pm$0.5 kpc, while for the Y2 region the size is $\sim$0$\farcs$5$\pm$0$\farcs$1, corresponding to $\sim$2.4$\pm$0.5 kpc. 
For region 3 we measure an angular size of $\sim$0$\farcs$3 or 1.4 kpc, all of them uncorrected for lensing magnification. In order to correct for the distorsions of the size measurements caused by the lensing effects, we used the HST images. This is done by comparing the physical sizes of different images of the same source caused by lensing. From the HST data presented by \citet{swinbank11} we conclude that the spatial magnification corresponds to a factor of $\sim$2. Hence, applying this correction factor, we find that the Y1 Region has an intrinsic size of $\sim$1.4 kpc, while for the Y2 Region the size is $\sim$1.2 kpc and 0.7 kpc for region Y3.

\subsubsection{Offset H$\alpha$ Emission}

Sources J033246, J033249, L50879 and region 2 of SMMJ2135-0102 show a clear spatial displacement between the H$\alpha$ and the continuum emission, as can be seen in
Figure~\ref{source_images}. These spatial offsets range from $\sim$0$\farcs$13 to 0$\farcs$50, corresponding to $\sim$1.2-4.2 kpc, relative to the center of the base 
continuum emission at nearby wavelengths. These offsets could be explained by off-nuclear star formation, perhaps as a consequence of a recent major merger similar to the one found in II Zw 096, 
where 80\% of the total infrared luminosity comes from an extremely compact, red, source not associated with either nucleus of the merging galaxies \citep{inami10}. Alternately, the offset H$\alpha$ distribution 
could result from a superwind blowing out of the galaxy, similar to the one detected in M82 \citep{lennart99}. In the latter scenario, the asymmetry would be due to obscuration of the receding wind by the galaxy, 
as a consequence of a viewing angle which is not edge-on. Furthermore, the H$\alpha$ offset could be explained by the effects of obscuration in the direction of the nucleus. Recently, \citet{chen15} also find 
significant displacements between the positions of the H$_{160}$-band continuum and the 870 $\mu$m emission in SMGs. This suggests that the dusty starburst regions and the less-obscured stellar distributions 
are not co-located.

\subsection{Extracted Integrated 1-D Spectra}
\label{sec:1dspec}

\begin{deluxetable*}{lccccc}
\tablewidth{0pt} 
\tablecaption{Spectroscopic measurements derived from the VLT/SINFONI data.\label{tab_spec_meas}}
\tablecolumns{6}
\tablehead{ 
\colhead{Name} &
\colhead{$z$\tablenotemark{a}} &
\colhead{Flux$_{H\alpha}$\tablenotemark{b}} &
\colhead{EQW$_{H\alpha}$ } &
\colhead{FWHM$_{H\alpha}$} &
\colhead{[NII]/H$\alpha$} \\
\colhead{} &
\colhead{} &
\colhead{[10$^{-17}$ergs$^{-1}$cm$^{-2}$]} &
\colhead{[\AA$_{rest}$]} &
\colhead{[km~s$^{-1}$]} &
\colhead{} 
}
\startdata 
 J033129\tablenotemark{c} & 1.482 & $<$45 & -- & -- & --\\
 J033212\tablenotemark{c} & 1.93   & $<$39 & -- & -- & --\\
 J033246 & 1.383 &27$\pm$4 & 45$\pm$8 & 106$\pm$38 &0.128$\pm$0.01\\
 
 J033249 & 2.327 &42$\pm$5 & 56$\pm$9 & 300$\pm$30&0.144$\pm$0.02\\
 \hspace{12pt}  Region 1 & 2.327 & 17$\pm$2 & 20$\pm$2 & 211$\pm$39 & -- \\
\hspace{12pt}  Region 2 &	 2.327 & 13$\pm$4 & 29$\pm$3 & 218$\pm$24 & --\\
 
 L50879 & 2.509 & 29$\pm$2 & 42$\pm$5 & 88$\pm$14 & -- \\
 
 RGJ0302+0010 & 2.236 & 54$\pm$7 & 41$\pm$9 & 173$\pm$59 &0.80$\pm$0.40 \\
\hspace{12pt}   Region 1 & -- & 23$\pm$8 & 42$\pm$10 & 103$\pm$29 & 0.71$\pm$0.13 \\
\hspace{12pt}   Region 2 &-- & 25$\pm$5 & 47$\pm$8 & 144$\pm$15 & 0.13$\pm$0.03  \\
 
 SMMJ04431+0210 (N4) & 2.509 &18$\pm$5 & 46$\pm$2 & 430$\pm$29 &0.47$\pm$0.06\\ 
\hspace{12pt} 	Region 1  & -- &7$\pm$1 & 112$\pm$15 & 397$\pm$57 &0.76$\pm$0.3 \\
\hspace{12pt} 	Region 2  & -- &5$\pm$1 & 30$\pm$4 & 175$\pm$27 &0.06$\pm$0.10  \\
 
 SMMJ2135-0120 & 2.323 & 24$\pm$2 & 62$\pm$4 & 127$\pm$16 &0.15$\pm$0.02 \\
\hspace{12pt}         Region 1  & -- & 10$\pm$2& 62$\pm$4 & 127$\pm$16 &0.15$\pm$0.02 \\
\hspace{12pt}         Region 2  & -- & 15$\pm$3 & 76$\pm$11 & 85$\pm$26 &0.18$\pm$0.10\\
\hspace{12pt}         Region 3  & -- & 6$\pm$1 & 24$\pm$2 & 48$\pm$4.7 &0.13$\pm$0.10 
\enddata
\textbf{
\tablenotetext{a}{Typical redshift uncertainty for the sources detected by SINFONI is $\sim$0.001.}
\tablenotetext{b}{Observed (i.e., not corrected for extinction) H$\alpha$ flux measured from the 1D spectra.}
\tablenotetext{c}{3-$\sigma$ upper limits to the $H$$\alpha$ fluxes were determined from the noise properties in the measured continuum around the expected positions of the lines
based on the redshifts reported by \citet{casey11a}.}
}
\end{deluxetable*}

\begin{figure*}
\centering
\subfigure{\includegraphics[width=.49\textwidth]{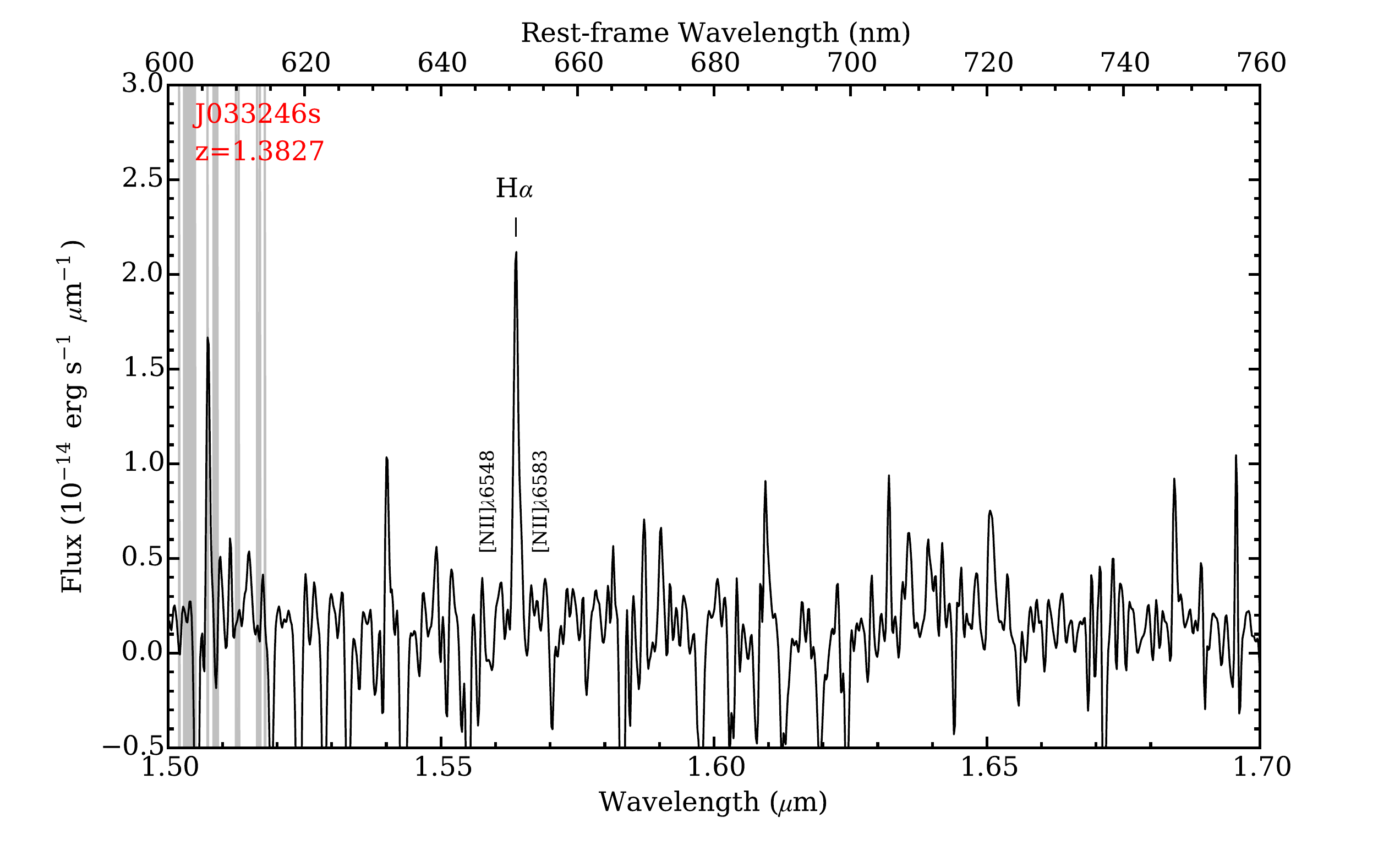}\label{J033246_1D}}
\subfigure{\includegraphics[width=.49\textwidth]{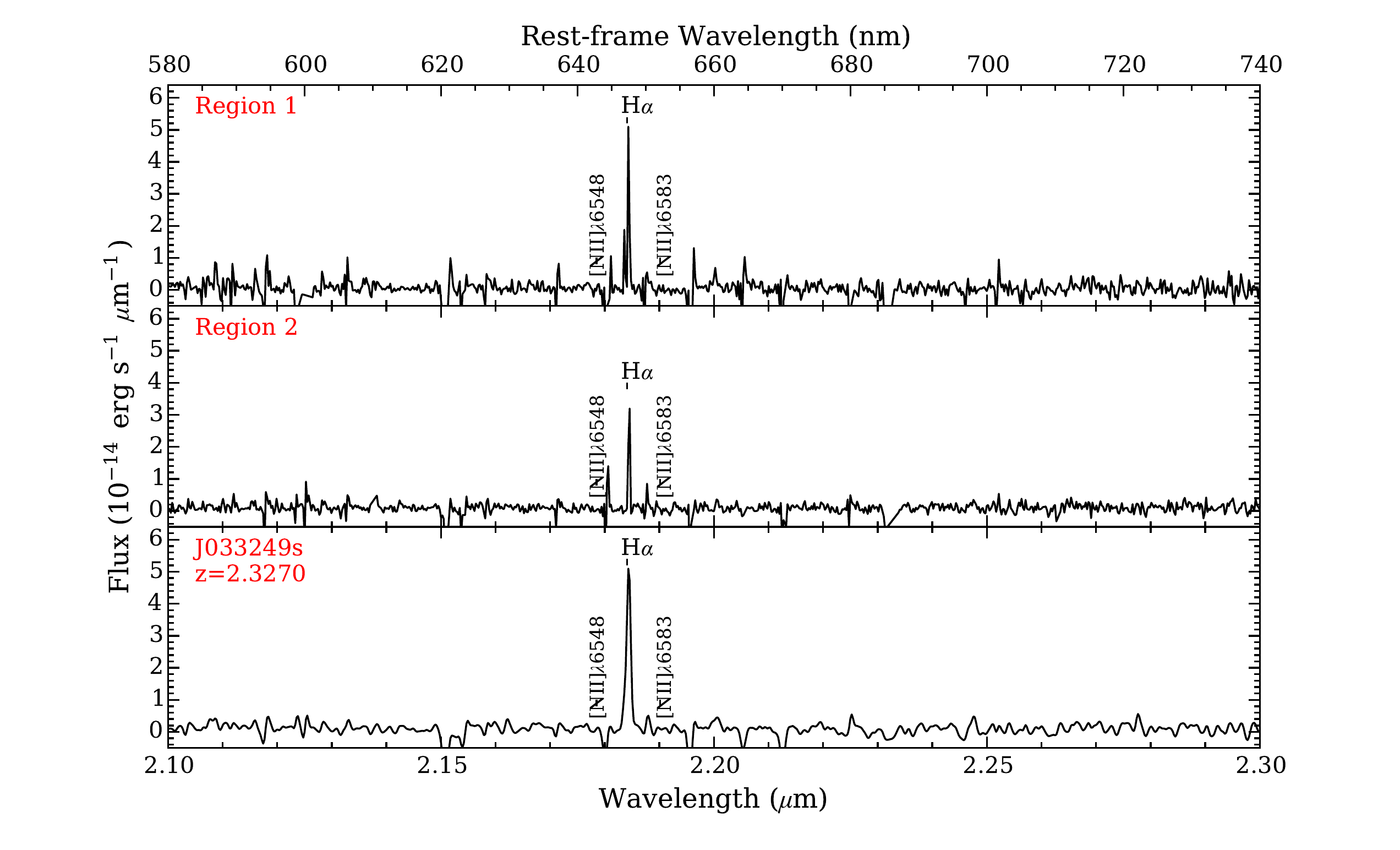}\label{J033249_1D}}\\
\subfigure{\includegraphics[width=.49\textwidth]{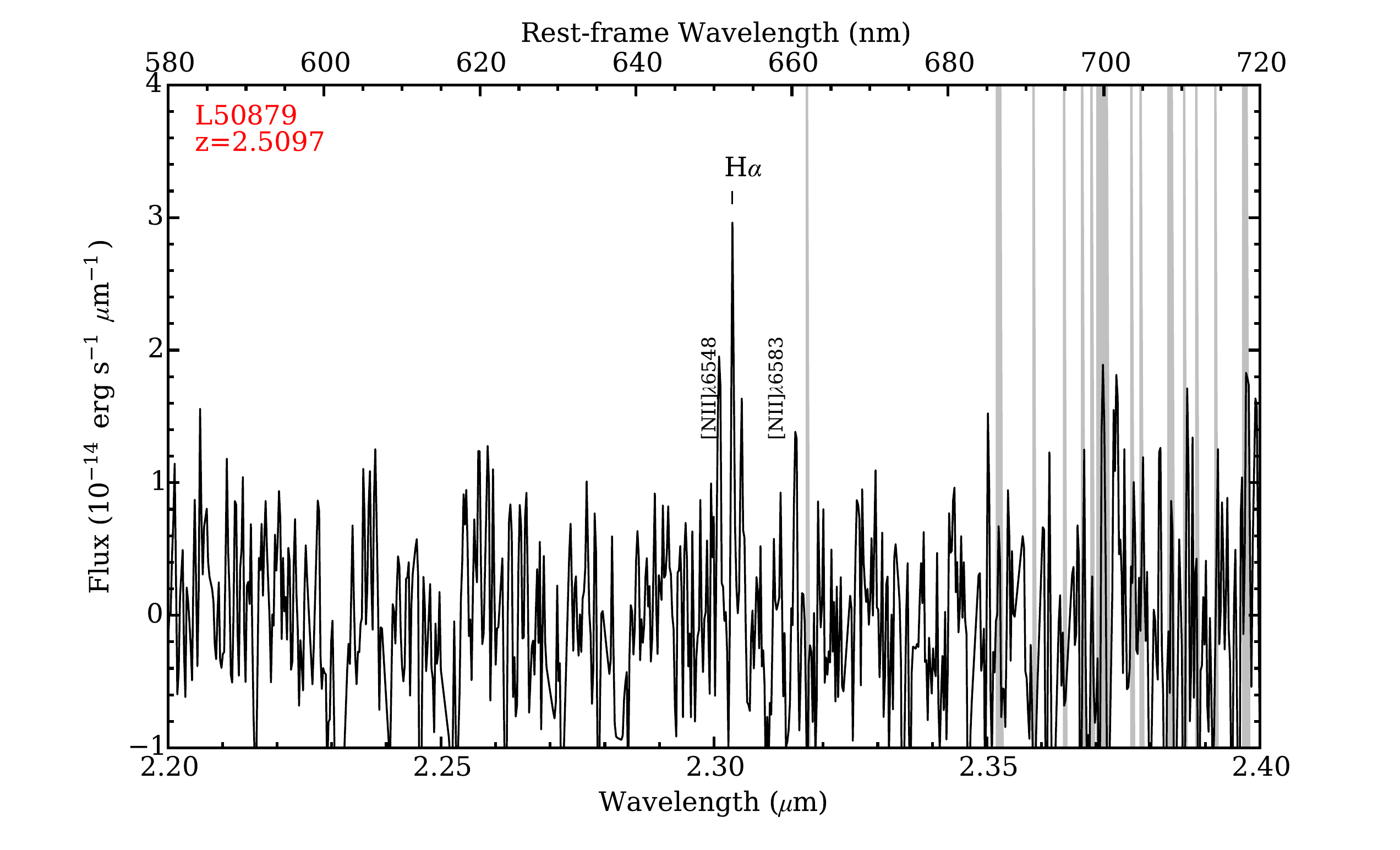}\label{L50879_1D}}
\subfigure{\includegraphics[width=.49\textwidth]{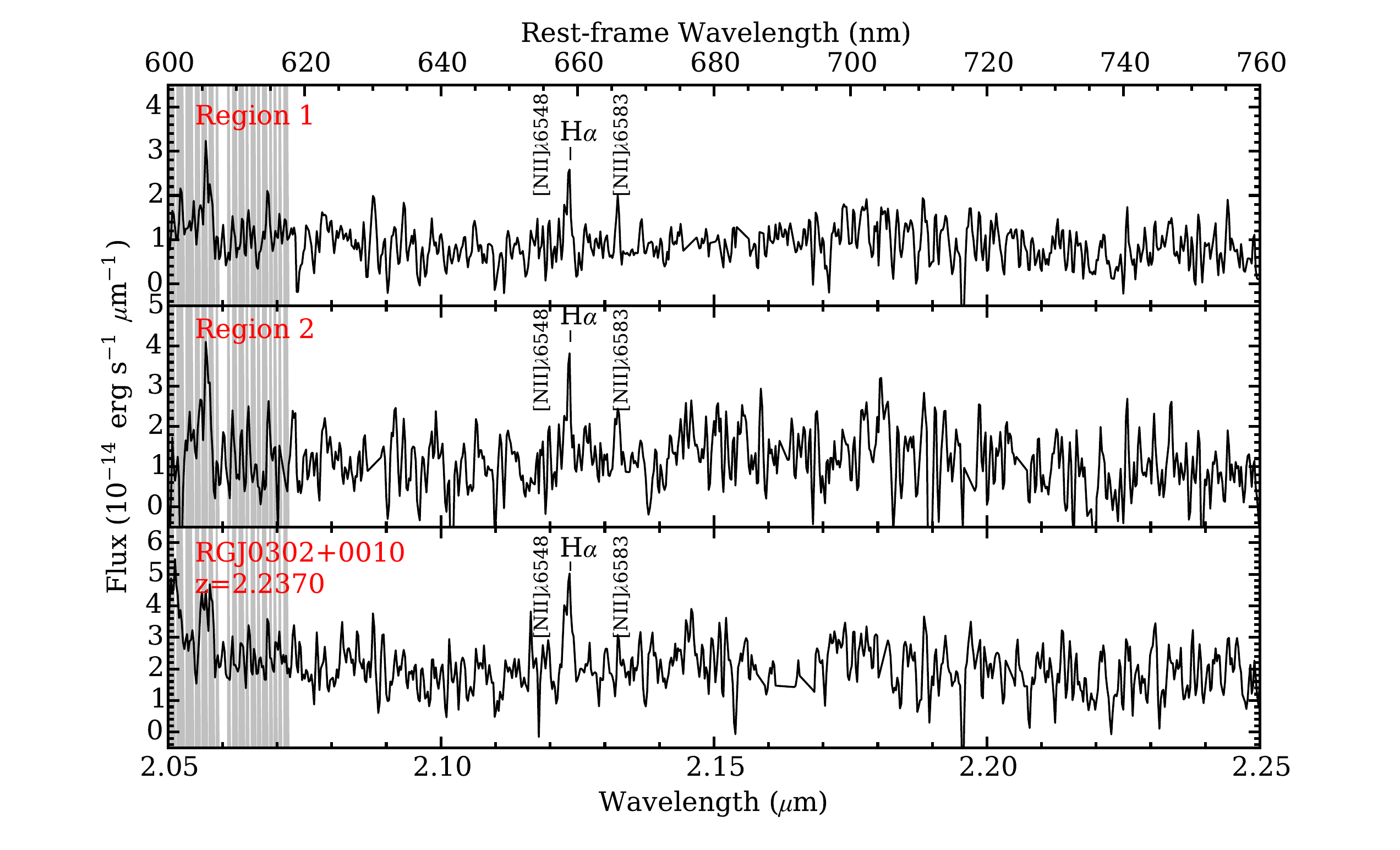}\label{J0302_1D}}\\
\subfigure{\includegraphics[width=.49\textwidth]{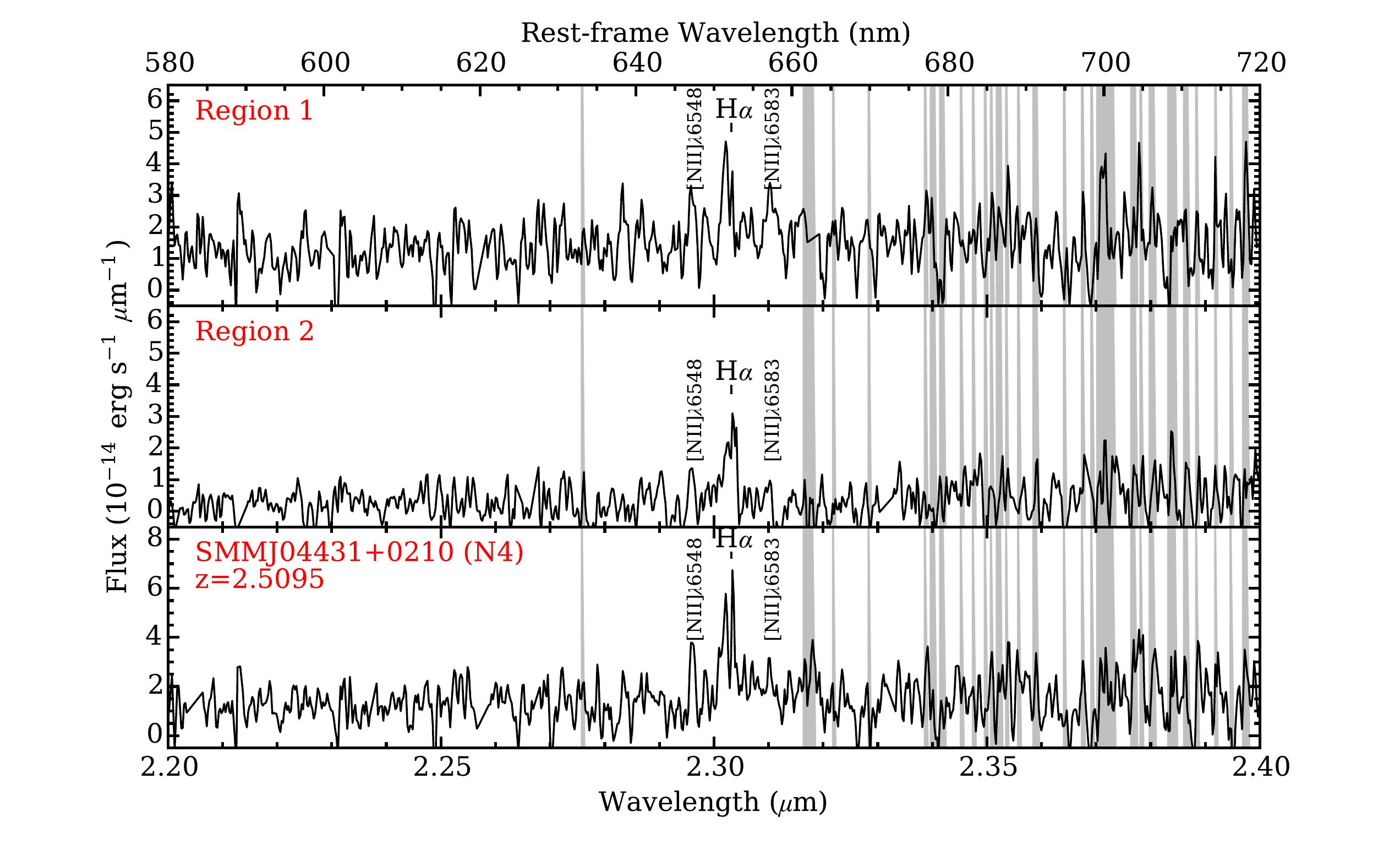}\label{smmj04431_1d}}
\subfigure{\includegraphics[width=.49\textwidth]{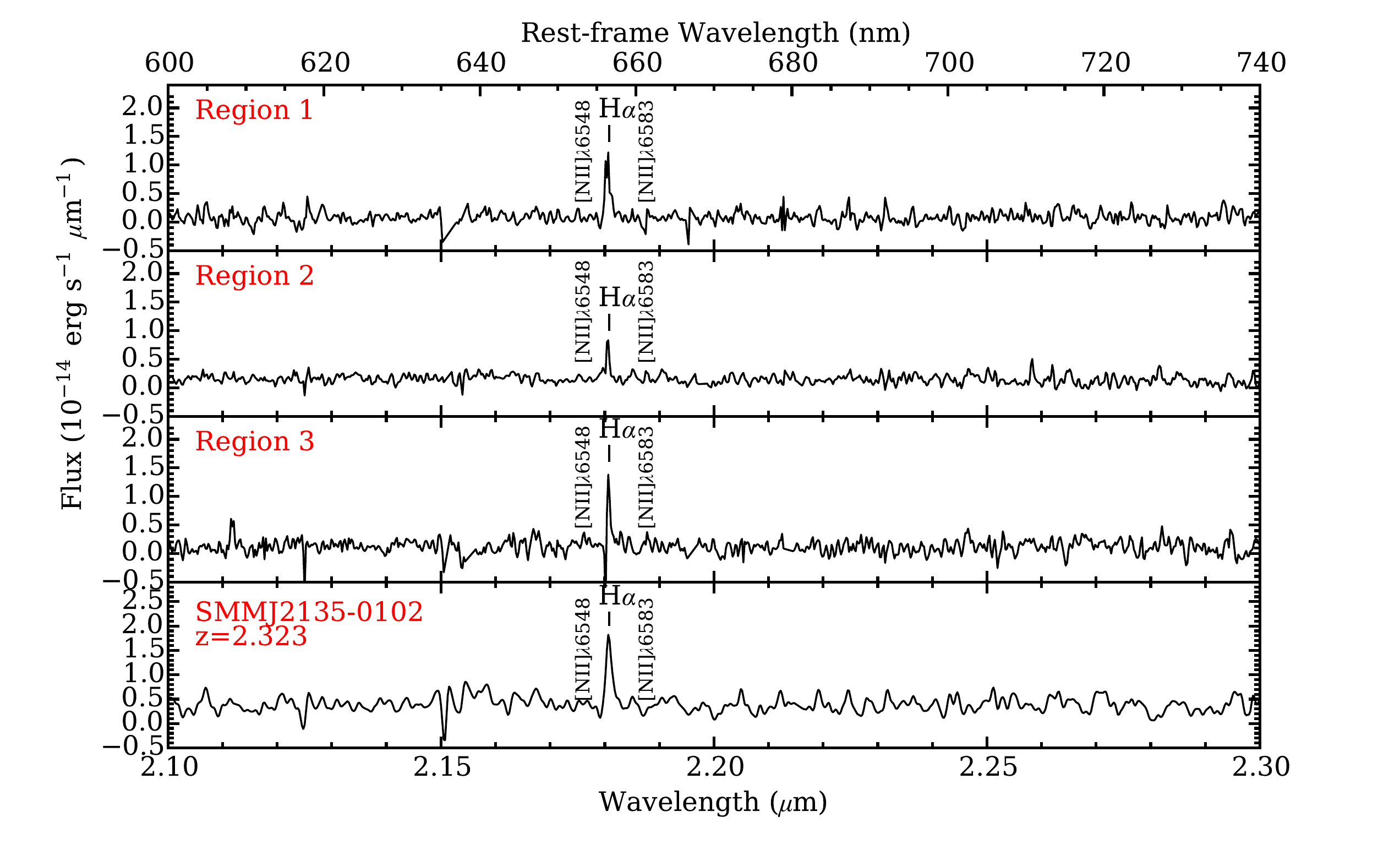}\label{smmj2135_1d}}
\caption{Extracted VLT SINFONI spectra for the SMGs in our sample. The gray zones represent the sky lines. The vertical solid lines indicate the H$\alpha$, [NII]$\lambda$6583 and [NII]$\lambda$6548 emission lines 
at redshifts of $z$=1.383 for J033246, $z$=2.327 for J03349, $z$=2.5097 for L50879, $z$=2.236 for RGJ0302+0010, $z$=2.5095 for SMMJ04431+0210 (N4) and $z$=2.3230 for SMMJ2135-0102.
\label{fig_1dspec}}
\end{figure*}

Figure~\ref{fig_1dspec} shows the extracted 1-D spectra for the six sources in our sample detected by our SINFONI observations. The spectra were obtained using the pseudo-longslits defined in the 
source images shown in Figure~\ref{source_images}. The redshift of each source was confirmed by the detection of the H$\alpha$ emission line. For the undetected sources, J033219 and J033212 
we derived the 3$\sigma$ H$\alpha$ flux upper limits by measuring the noise properties of the continuum regions surrounding the expected wavelengths of the lines based on the redshift values 
provided by \citet{casey11a}. 

For the detected sources, we then used these spectra to classify each spatially-resolved region as dominated by either AGN or star-formation using the classification scheme based on emission line 
ratios of \citet{kewley06}. Ideally, this is done using a combination of line ratios such as [NII]/H$\alpha$ and [OIII]/H$\beta$. However, in this case only the H$\alpha$ and [NII] emission lines are 
available in our SINFONI data. Hence, we adopt a [NII]/H$\alpha$$>$0.7 flux ratio for sources or regions classified as AGN dominated, based on the \citet{kewley06} classification scheme.

\paragraph{J033246} 

From the SINFONI H-band spectrum of J033246 we derive a redshift of $z$=1.383 based on the H$\alpha$ and [NII]$\lambda$6583 lines, fully consistent with the value reported by \citet{casey11a} of 
$z$=1.382 from VLT/ISAAC observations. The intrinsic H$\alpha$ line width is 130$\pm$38~km~s$^{-1}$, in line with the value previously reported by \citet{casey11a} of 150$\pm$70~kms$^{-1}$. The 
rest-frame [NII]/H$\alpha$ line flux ratio is 0.13$\pm$0.01, also in accord with the value found by \citet{casey11a}, which suggests that J033246 is consistent with star-forming activity. 
Starting from the measured spatially-integrated H$\alpha$ line flux, and assuming that it is entirely due to star formation, we can estimate the star formation rate for J033246 using the relation reported by \citet{kennicutt98}:

\begin{equation}
SFR [M_{\odot} yr^{-1}] = 7.9 \times 10^{-42} L(H\alpha) [erg s^{-1}].
\end{equation}

We find a SFR of 26$\pm$4~M$_{\odot}$yr$^{-1}$. Using the measured size and derived SFR, we compute a star formation rate surface of $\Sigma_{SFR}$=0.5~M$_{\odot}$yr$^{-1}$kpc$^{-2}$.

\paragraph{J033249}
For J033249 we measure a redshift of $z$=2.327 based on the H$\alpha$ and [NII]$\lambda$6583 lines. The rest frame [NII]/H$\alpha$ line flux ratio is  0.144$\pm$0.02, fully consistent with the value reported by
\citet{casey11a} of 0.14$\pm$0.10, while for H$\alpha$ we find a FWHM of 310$\pm$30~km~s$^{-1}$, consistent with the value of 360$\pm$50~km~s$^{-1}$ reported by \citet{casey11a}. 
From the observed H$\alpha$ flux we derive a SFR of 132$\pm$17~M$_{\odot}$yr$^{-1}$. Using these estimates, we compute a star formation rate surface density of 
$\Sigma_{SFR}$=7.9~M$_{\odot}$yr$^{-1}$kpc$^{-2}$.

\paragraph{L50879}
In the case of L50879, we measure a redshift of 2.5097 from the H$\alpha$ and [NII]$\lambda$6583 lines. The measured rest-frame [NII]/H$\alpha$ line flux ratio is 0.3$\pm$0.7,
suggesting that L50879 is dominated by star-forming activity. The intrinsic H$\alpha$ line width is 42$\pm$5~km~s$^{-1}$. From the integrated H$\alpha$ emission of the galaxy 
we derive a SFR of 60$\pm$8~M$_{\odot}$yr$^{-1}$. We then compute a star formation rate surface density of $\Sigma_{SFR}$=2.4~M$_{\odot}$yr$^{-1}$kpc$^{-2}$.

\paragraph{RGJ0302+0010}
For RGJ0302+0010 we derive a redshift of $z$=2.237 based on the H$\alpha$ and NII$\lambda$6583 lines. Similar results were reported previously using Keck/NIRSPEC and VLT/ISAAC \citep{swinbank04} 
and NIFS at Gemini-North \citep{harrison12}. The intrinsic FWHM of H$\alpha$ is 188$\pm$59 km~s$^{-1}$. The rest frame [NII]/H$\alpha$ line flux ratio is measured to be 0.8$\pm$0.4, marginally suggesting 
the presence of AGN activity. This is consistent with the values found by \citet{chapman04} and \citet{swinbank04} of 1.1$\pm$0.4. 
Similarly, the value of $\log$(O[III]/H$\beta$)=0.97$\pm$0.13 (O[III]/H$\beta$=9.3$\pm$1.3) previously found by \citet{harrison12}, further confirms this conclusion regarding the AGN activity in this galaxy. 
From the galaxy-wide integrated H$\alpha$ emission we derive an SFR of 136$\pm$20~M$_{\odot}$yr$^{-1}$, assuming that the H$\alpha$ is only due to star formation, which as we discussed above is unlikely
to be the case. Region 1 has  [NII]/H$\alpha$=0.71$\pm$0.13 and a line ratio [O III]/H$\beta$=9$\pm$3 indicating that this this is the region where the AGN is located, while Region 2 has 
[NII]/H$\alpha$=0.13$\pm$0.03, likely suggesting that it is dominated by star-forming activity. The SFR derived from the H$\alpha$ line emission is 75$\pm$20~M$_{\odot}$yr$^{-1}$ for Region 1, and 
105$\pm$14~M$_{\odot}$yr$^{-1}$ for Region 2. Using these estimates of the sizes of H$\alpha$ and SFR derived from the H$\alpha$ emission, we compute the star formation rate surface densities and 
find $\Sigma_{SFR}$=17~M$_{\odot}$yr$^{-1}$kpc$^{-2}$ for Region 1 and  $\Sigma_{SFR}$=5.7~M$_{\odot}$yr$^{-1}$kpc$^{-2}$ for Region 2.

\paragraph{SMMJ04431+0210 (N4)} 
For SMMJ04431+0210 (N4) we derive a redshift of $z$=2.5095 based on the H$\alpha$ and [NII]$\lambda$6583 lines, similar to the value found by \citet{frayer03} of $z$=2.5092 from Keck/NIRSPEC observations.
The SINFONI K-band spectrum, presented in Fig.~\ref{smmj04431_1d}, shows a strong H$\alpha$ line with a FWHM of 437$\pm$29 km~s$^{-1}$, somewhat smaller than the value reported by \citet{frayer03} of 
520$\pm$40 km~s$^{-1}$, and larger than the measurement by \citet{neri03} of 350$\pm$60 km~s$^{-1}$. The observed [NII]$\lambda$6583 line is narrower, having an intrinsic FWHM of 165$\pm$28 km~s$^{-1}$. 
The observed [NII]$\lambda$6583/H$\alpha$ flux ratio is 0.47$\pm$0.06. 

We find a SFR of 77$\pm$20~M$_{\odot}$yr$^{-1}$ (uncorrected for lensing magnification), and 17.5$\pm$4.8~M$_{\odot}$yr$^{-1}$ corrected for lensing magnification, assuming an amplification 
factor of $\mu$=4.4 as measured by \citet{smail99}.  We find that region 1 is consistent with ionization due to an AGN, given the observed flux ratio of [NII]/H$\alpha$=0.76$\pm$0.3 and 
a SFR of  30$\pm$3.7~M$_{\odot}$yr$^{-1}$, while region 2 is probably dominated by star formation processes, as it has a flux ratio of [NII]/H$\alpha$=0.06$\pm$0.1 and SFR of 20$\pm$2.9~M$_{\odot}$yr$^{-1}$, 
all of them uncorrected for lensing magnification. Combining the H$\alpha$ SFR with the observed sizes of each region we can compute SFR surface densities of 
$\Sigma$$_{SFR}$ of 2.3~M$_{\odot}$yr$^{-1}$kpc$^{-2}$ for the entire galaxy,  $\Sigma$$_{SFR}$=2.4~M$_{\odot}$yr$^{-1}$kpc$^{-2}$ for region 1 and
$\Sigma$$_{SFR}$=7.9~M$_{\odot}$yr$^{-1}$kpc$^{-2}$ for region 2.

\paragraph{SMMJ2135-0102} 
Figure~\ref{smmj2135_1d} shows the extracted one-dimensional K-band spectrum for each of the three regions in SMMJ2135 and the galaxy-wide integrated spectrum. We confirm the redshift of the source at
$z$=2.323 based on the H$\alpha$ and [NII]$\lambda$6583 lines, consistent with the value found by \citep{swinbank10} of $z$=2.3259 based on  GBT/Zpectrometer Observations of CO lines.
The overall [NII]/H$\alpha$ line flux ratio is found to be 0.15$\pm$0.02 suggesting that the observed emission lines are dominated by star formation activity. 
From the integrated H$\alpha$ emission we derive a SFR of 82$\pm$4~M$_{\odot}$yr$^{-1}$, uncorrected for lensing magnification and extinction. 
We measure star formation rates of 33$\pm$7~M$_{\odot}$yr$^{-1}$, 54$\pm$8~M$_{\odot}$yr$^{-1}$ and 20$\pm$3~M$_{\odot}$yr$^{-1}$ for regions Y1, Y2 and 3 respectively.
Using these estimates of the sizes of H$\alpha$ present in the previous section and SFR from H$\alpha$ emission for these regions we derive star formation rate surface of 
$\Sigma_{SFR}$=16.8~M$_{\odot}$yr$^{-1}$kpc$^{-2}$ for region Y1, 
$\Sigma_{SFR}$=37.5~M$_{\odot}$yr$^{-1}$kpc$^{-2}$ for region Y2, and  
$\Sigma_{SFR}$=40.8~M$_{\odot}$yr$^{-1}$kpc$^{-2}$ for region 3.

\subsection{Velocity Maps}
\label{sec:vmap}
\begin{figure*}
\centering
\subfigure{\includegraphics[width=.99\textwidth]{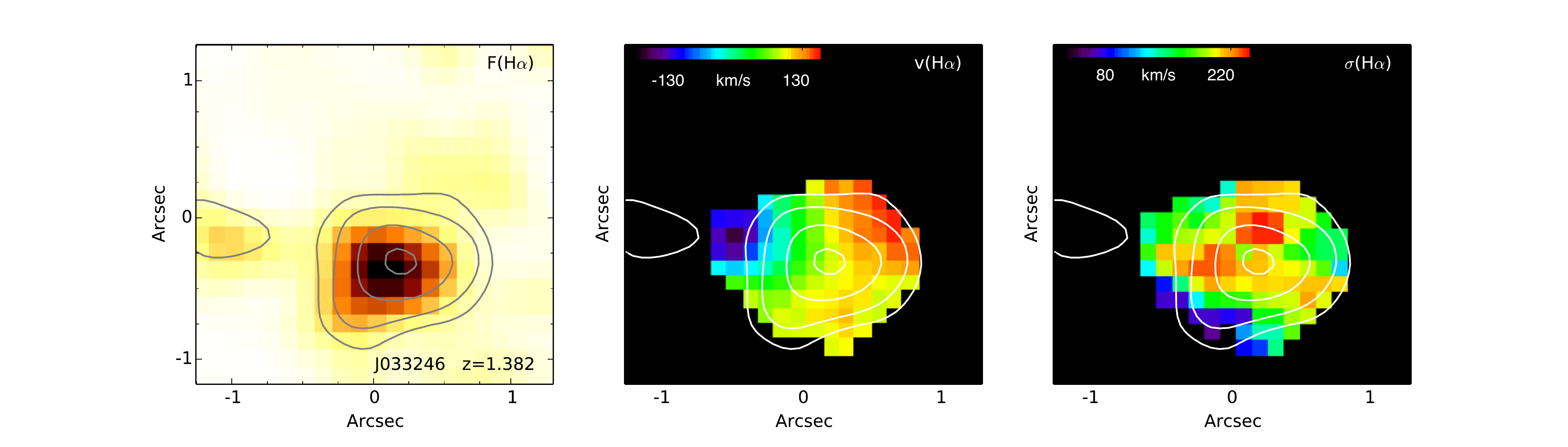}}
\subfigure{\includegraphics[width=.99\textwidth]{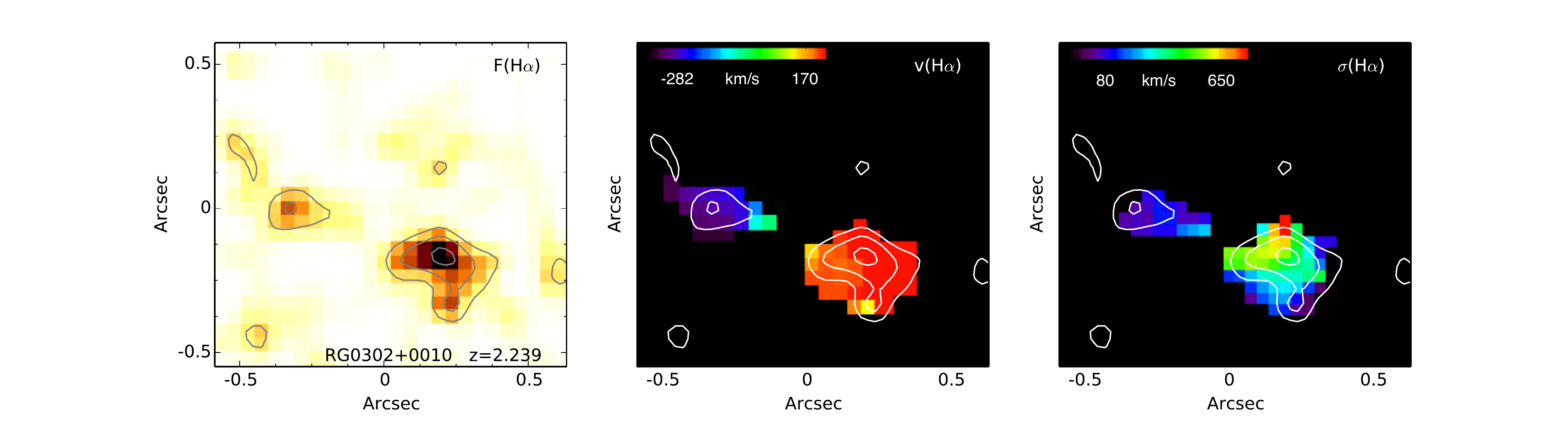}}
\subfigure{\includegraphics[width=.99\textwidth]{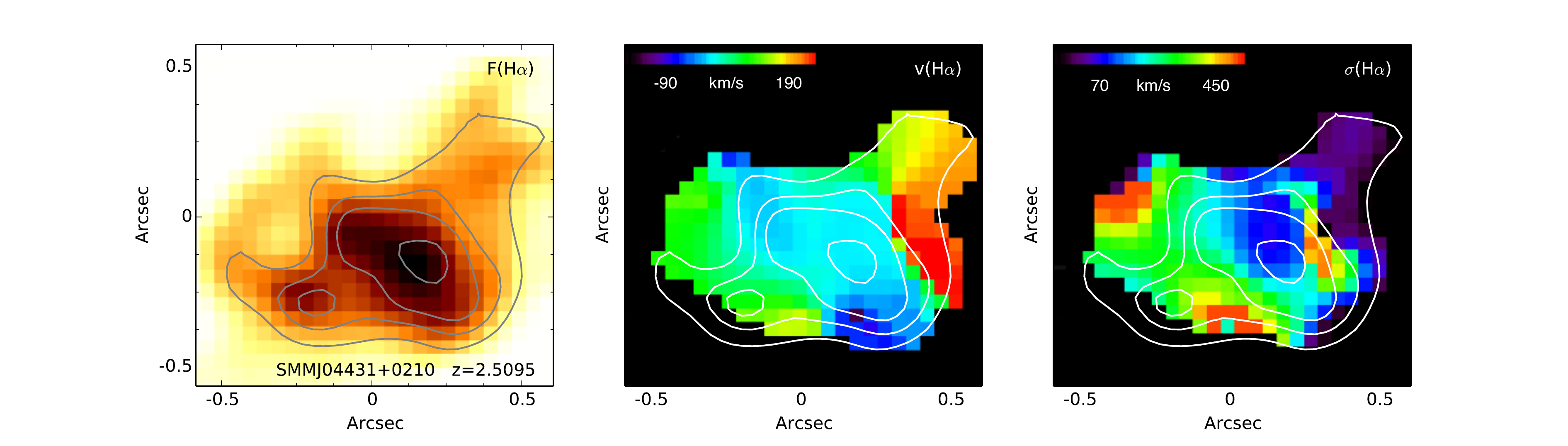}}
\subfigure{\includegraphics[width=.99\textwidth]{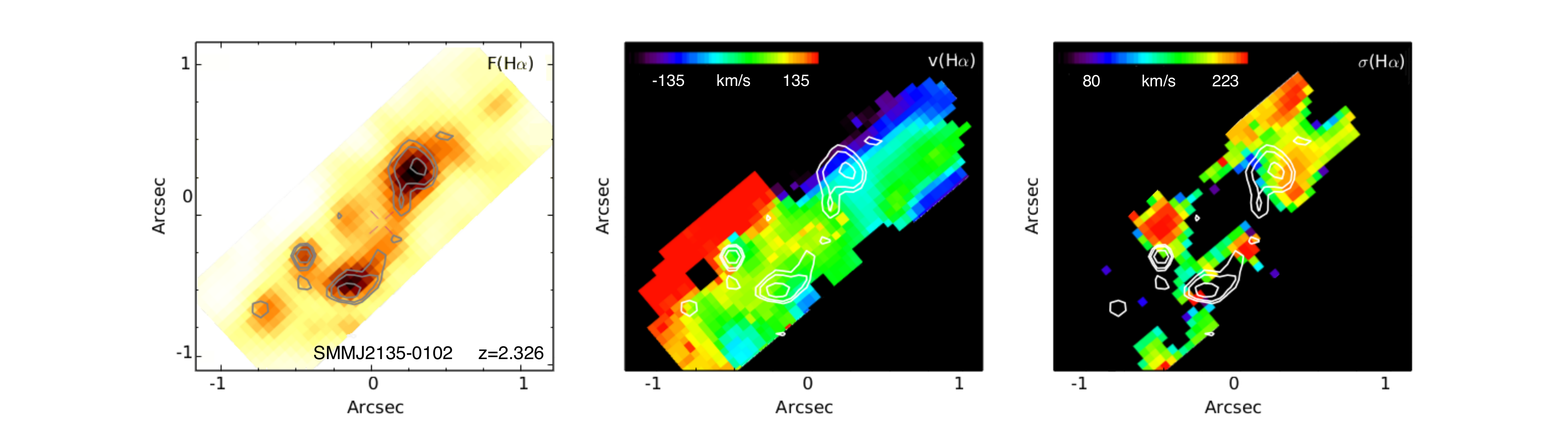}}
\caption{Maps of the H$\alpha$ properties for four of the sources in our sample. Top to bottom: J033246, RGJ0302+0010, SMMJ04431+0210 (N4) and, SMMJ2135-0102. {\it Left panel}: H$\alpha$ intensity 
contours in gray overlapped on the VLT/SINFONI continuum maps. {\it Middle panel}: H$\alpha$ velocity field, in km s$^{-1}$. {\it Right panel}: H$\alpha$ velocity dispersion field, in km s$^{-1}$.\label{fig_velmap}}
\end{figure*}

In addition to the identification and redshift measurements of the galaxy responsible for the sub-mm emission, the VLT/SINFONI data cubes can be also used to study their dynamical properties.Velocity fields and 
velocity dispersion maps are measured from the shift in observed wavelength and the width of the H$\alpha$ line across each source. Figure~\ref{fig_velmap} shows the H$\alpha$ velocity maps for four of our targets:  
J033246, RGJ0302+0010, SMMJ04431+0210 (N4),  and SMMJ2135-0102. For the two remaining sources, J033249 and L50879, the signal to noise ratio of the H$\alpha$ line was not high enough to produce a velocity map.

J033246 has the most symmetric velocity field, showing some evidence of rotation, with a gradient of $\sim$250~km/s from one side to the opposite. SMMJ04431+0210 (N4) is 
clearly irregular, both in the flux distribution and velocity profile, with no evidence of ordered motions. Finally, both RGJ0302+0010 and SMMJ2135-0102 present a clumpy structure, with no signs of ordered motions.

In about half of the targets in our sample, J033246, RGJ0302+0010, SMMJ04431+0210 (N4) and SMMJ2135-0102, we find velocity offsets of $\sim$ few $\times$100 km~s$^{-1}$ between distinct galactic-scale 
regions, with irregular kinematics. These velocity offsets could be explained by invoking a merger scenario, as presented by e.g., \citet{engel10}. This is also supported by the clearly disrupted morphologies shown by 
the deep \textit{HST} images available for these sources. Evidence for a connection between high luminosity SMGs and major mergers has been previously and extensively reported in the literature \citep[e.g.,][]{smail98,smail04,ivison10,swinbank10,alaghband-zadeh12}.

\section{Discussion}
\label{sec:discussion}

\subsection{H$\alpha$ Morphologies}

We have spatially mapped the H$\alpha$ line emission for a sample of six high-redshift luminous SMGs. As presented in Section~3.1, the H$\alpha$ morphologies in our sample tend to be irregular 
and/or clumpy. We measure the H$\alpha$ emission to be extended on scales $>$4 kpc, with an average H$\alpha$ half-light radius of $<$r$_{1/2}$$>$=3.2$\pm$1.8~kpc. These sizes are in 
good agreement with previous, seeing-limited measurements of the extent of SMGs in H$\alpha$ \citep[4--16 kpc;][]{menendez-delmestre13,alaghband-zadeh12} and nebular lines 
\citep[4--11 kpc;][]{swinbank06}. Similarly, recent high-resolution measurements (0.3$''$) using the Atacama Large Millimeter Array (ALMA) 870 $\mu$m of 52 bright SMGs in the Ultra Deep Survey 
(UDS) field reported by \citet{simpson15a} yielded a median physical half-light diameter of 2.4$\pm$0.2 kpc. However using HST/WFC3, \citet{chen15} reported a median half-light radius of 
4.4$^{+1.1}_{-0.5}$ kpc for a sample of 48 bright SMGs in the Extended Chandra Deep Field South (ECDF-S). Our H$\alpha$ size measurements are also consistent with the expectations based
on theoretical simulations of mergers with high gas fractions (e.g., \citealp{mihos99,narayanan10}).

For three galaxies in our sample, we were able to resolve the H$\alpha$ emission into two or more galactic-scale extensions/components on $\sim$0.5-1.5 arcsec (3-11 kpc) scales. In two of these 
three systems we were able to identify possible signs of AGN activity based on the observed [NII]/H$\alpha$ flux ratio. Previous IFU studies of SMGs have targeted galaxies showing AGN signatures 
in their H$\alpha$ spectra \citep{menendez-delmestre13}, while just two systems in our sample display clear H$\alpha$ AGN signatures.

In the remaining sources in our sample we find undisturbed morphologies, with extended spatial extensions (size $\sim$ 8-11 kpc) of narrow-line H$\alpha$ emission that contribute a significant 
fraction ($\sim$40\%-90\%) of the galaxy-wide H$\alpha$ emission. Furthermore, we also find a lower [NII]/H$\alpha$ flux ratio, thus strongly suggesting that the H$\alpha$ component in these systems 
is associated with star-forming activity. High-resolution radio continuum observations \citep{chapman04,biggsyivison08} found spatial extensions of $\sim$8-10 kpc in diameter, while a high-resolution 
far-IR study by \citet{younger10} revealed that this emission typically extends out to spatial scales in the range of $\sim$5-8 kpc. In high-resolution observations of a range of CO transitions 
\citep[e.g.,][]{engel10,ivison11,tacconi06,tacconi08,hainline06}, the size measurements of SMGs from CO fluxes are in the range of $\sim$1-16 kpc.

Our results indicate that the SMG clumps in our sample have high star formation surface densities  (Section \ref{sec:1dspec}), close to those found in local extreme environments, such as in 
circum-nuclear starbursts and luminous infrared galaxies. In some of our targets, the H$\alpha$ distribution appears to be offset from the observed-frame near-IR continuum emission traced by 
SINFONI and \textit{HST} observations. Considering the much greater spatial extents found for these SMGs ($\sim$ 3-11 kpc) in comparison to the $\sim$1-kpc sized nuclear starbursts 
\citep{kennicutt98} and the $\sim$100 pc starburst regions observed in local ULIRGs (e.g., \citealp{scoville00}), SMGs appear to be undergoing this intense activity over much larger spatial 
scales. This is well in line with similar conclusions reported by several studies in the past \citep{swinbank06,alaghband-zadeh12, menendez-delmestre13}.

\subsection{Kinemetry Analysis}

As described in Section \ref{sec:vmap}, the SMGs studied in this work present a wide range of H$\alpha$ kinematical properties. In order to establish if the motions observed in these galaxies are dominated by 
ordered (e.g., rotation) or random (as caused by e.g., a recent major galaxy merger) dynamics, we employ kinemetry analysis. This is done by measuring the level of symmetry in the mean of the velocity and 
the velocity dispersion fields. The kinemetry measurement procedure described by \citet{shapiro08} uses the method detailed in \citet{krajnovic06} to analyze maps of kinematic moments based on the line-of-sight-velocity 
distribution. This can be considered as an extension of the surface photometry to the higher-order moments of the velocity distribution. This technique was developed to study stellar dynamics of local ellipticals \citep{copin01}, 
but has been extended to the dynamics of high-$z$ galaxies \citep[e.g.,][]{forster-schreiber09}. More recently, \citet{alaghband-zadeh12} used a technique based on measuring 
velocity and velocity dispersion asymmetries in order to establish if galaxies are dominated by rotation or random motions. They applied this technique to SMGs similar to the ones in our sample.

Asymmetries in moment fields are measured by fitting ellipses, varying the position angle and ellipticity, to the velocity and dispersion fields. These coefficients, v$_{asym}$ and $\sigma$$_{asym}$, are used to establish the 
level of asymmetry. The two asymmetry measures are combined as K$_{asym}$ = $\sqrt{v_{asym}^{2} + \sigma_{asym}^{2}}$. Following the definition of \citet{alaghband-zadeh12}, systems with 
K$_{asym}$$>$0.5 are classified as mergers. Any asymmetries in these fields represent deviations from the idealized model (a rotating thin disk) and as such the combined asymmetry of the two fields 
can fully describe how accurately a system is represented by this idealized disk. This model displays an ordered velocity field, peaking at the semi-major axis of the ellipse and reaching a value of
zero at the semi-minor axis, and a centrally peaked velocity dispersion field. 

\begin{figure}
\centering
\includegraphics[width=.5\textwidth]{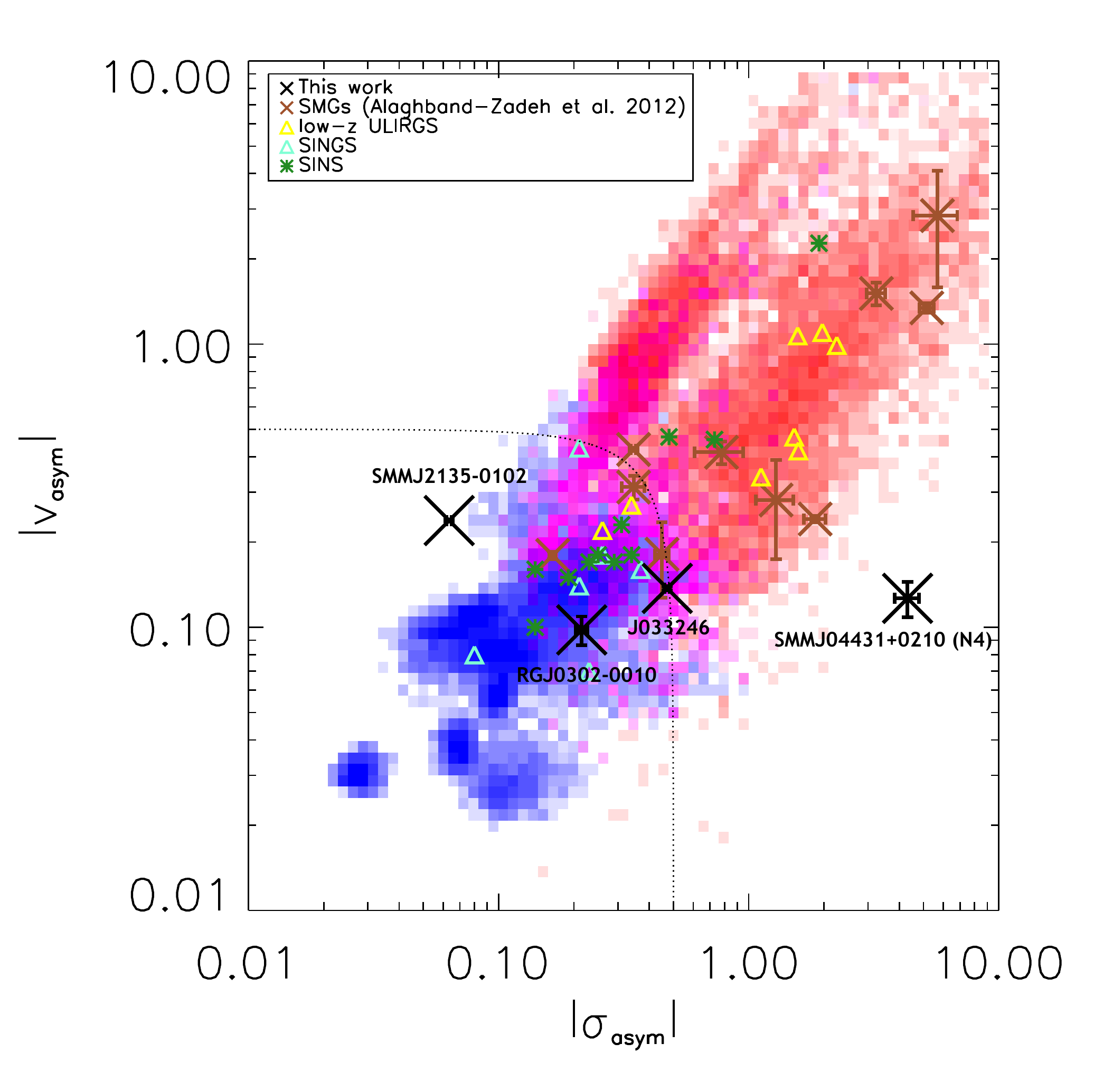}
\caption{Velocity asymmetry versus velocity dispersion asymmetry values for 4 SMGs of our sample: J033246, RGJ0302+0010, SMMJ04431+0210 and SMMJ2135-0102. These measurements 
were carried out using the procedure described by \citet{alaghband-zadeh12}. Values for the sources in our sample are compared against other groups of active galaxies such as local spiral galaxies 
in the SINGs sample, low-redshift ULIRGs and star-forming galaxies at $z$$\sim$2 in the SINS sample. The background blue points are template disks and red points are mergers, from the library 
of \citet{shapiro08}. While 3 out of the four sources, J033246, RGJ0302+0010 and SMMJ2135-0102 are located in the ``disk'' region of the diagram, only in one of them (J033246) did we find evidence 
for ordered motions. In the other two cases, these are clearly clumpy galaxies in which each individual clump is spatially unresolved. SMMJ04431+0210 (N4) is clearly in the ``mergers'' part of the 
diagram. \label{fig_kinemetry}}
\end{figure}

In Fig.~\ref{fig_kinemetry} we compare the levels of asymmetry in the velocity and velocity dispersion for the four SMGs in our sample for which velocity maps were obtained. As can be seen on that 
figure, and perhaps surprisingly, 3/4 of the galaxies in the SMG sample (J033246, RGJ0302+0010 and SMMJ2135-0102) are found in the region where disk-dominated sources are located. This is 
somewhat expected for J033246, since as discussed above, this source appears to be dominated by rotation. The other two sources appear to be clumpy in their H$\alpha$ morphologies, which might 
explain why their asymmetry values are lower than expected, as each individual clump is spatially unresolved and thus does not show significant asymmetries. Hence, their classifications based on the 
derived asymmetry values are probably explained by the limitations of the spatial resolution of our observations. Finally, SMMJ04431+0210 (N4) clearly shows very high asymmetry levels, as can be seen 
in Figure~\ref{fig_velmap}. Therefore, it is not surprising that this source is also classified as  ``merger-dominated'' in this diagram. Hence, the kinemetry analysis also suggests that most of these 
sources are dominated by random motions. This is in line with previous observations of similar high-luminosity IR and sub-mm galaxies at $z$$\sim$2, such as those reported by 
\citet{alaghband-zadeh12} and \citet{menendez-delmestre13} among others. A recent analysis presented by \citet{hung15}, which takes a sample of local IR-luminous galaxies and artificially 
redshifts their H$\alpha$ IFU data, shows that post-coalescence mergers may also display kinematics that can be wrongly classified as ``disk like'', as appears to be the case for some of our sources.

\subsection{H$\alpha$ Star Formation Rates}

We now compare the star formation rates for the galaxies in our sample, previously derived from far-IR observations, with those obtained from the H$\alpha$ observations described here. 
The implied star formation rates, including those derived from IR observations obtained from the literature are presented in Table~\ref{tab_spec_der}. 
Specifically, we use the values reported by \citet{casey11a} for J033246 and J033249 from Spitzer-MIPS, BLAST and LABOCA observations. For RGJ0302+0010 the values were derived by 
\citet{swinbank04} fitting model spectral energy distributions (SEDs) to the observed 850~$\mu$m and 1.4 GHz fluxes assuming that the local far-IR/radio correlation \citep{condon91, garrett02} holds 
at higher redshifts. For SMMJ04331+0210 we used the values obtained by \citet{neri03} from observed 850$\mu$m flux densities and assuming a modified grey-body model (T=40 K) and a 
frequency-dependent emissivity. Finally, for SMMJ2135-0120 values were derived by \citet{ivison10} from Herschel, SCUBA-2, VLA and APEX observations. Both SFR measurements for 
SMMJ2135-0120 were corrected for lensing magnification assuming a factor of $\mu$=32.5$\pm$4.5 \citep{swinbank10}, while a magnification factor of $\sim$=4.4 \citep{smail99} was assumed 
for SMMJ04431+0210 (N4). It should be noted that using the same lensing magnification for the H$\alpha$ and FIR fluxes is not ideal, as these emissions might come from slightly different regions 
and thus might present small variations in their magnifications.

\begin{figure}
\begin{center}
\includegraphics[width=0.5\textwidth]{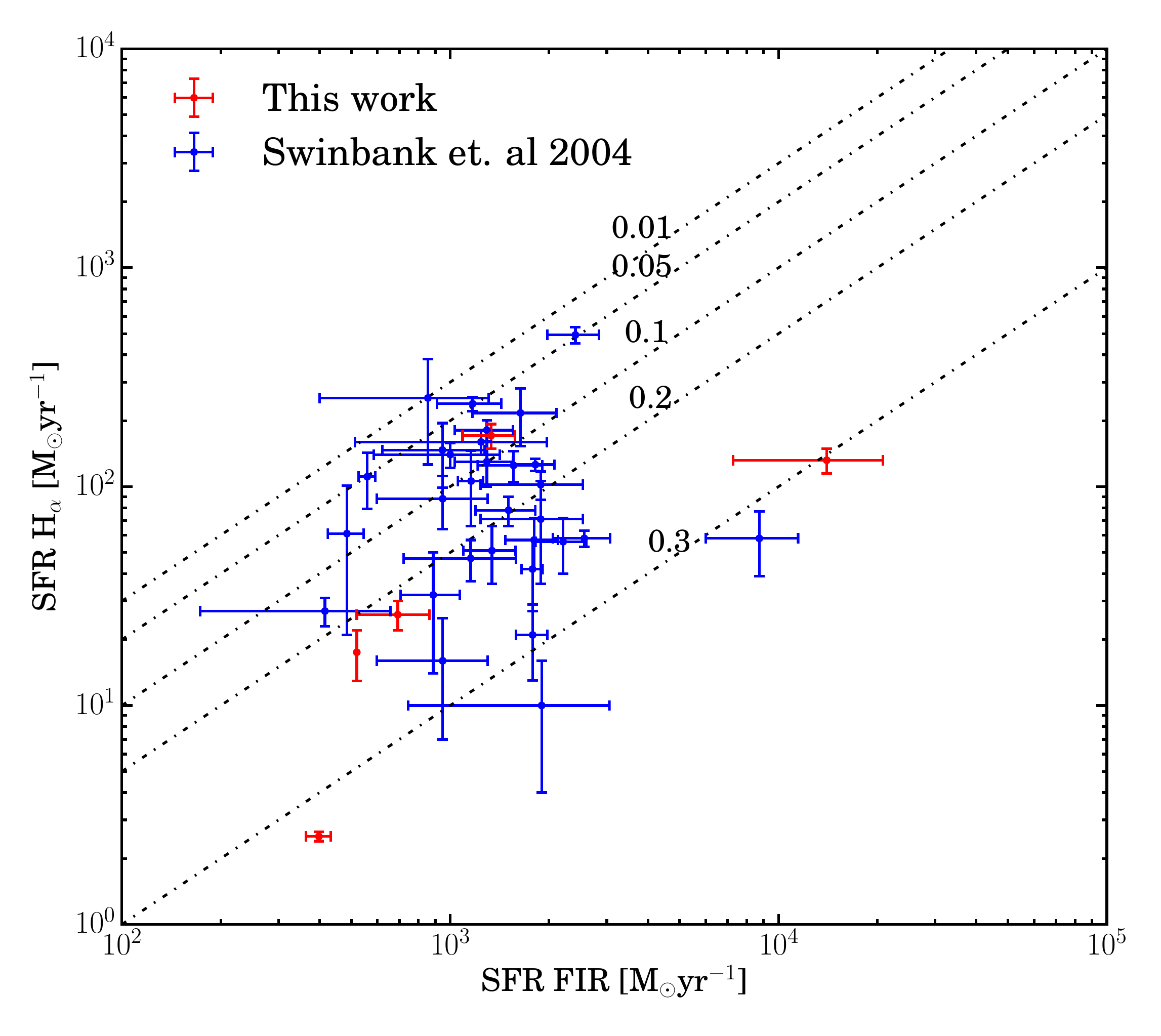}
\end{center}
 \caption{SFR(H$\alpha$) versus SFR(FIR). The {\it Red points} are measured values from the present work. SFR(H$\alpha$) values were derived from the H$\alpha$ luminosities obtained from our SINFONI 
 observations (not corrected for extinction). The FIR SFRs for J033246, J033249,  RGJ0302+0010, SMMJ04431+0210 (N4) and SMMJ2135-0210 were derived from FIR luminosities in the literature using the 
 Kennicutt relations \citep{kennicutt98}. For SMMJ04431+0210 (N4) the SFR values were corrected for lensing magnification assuming a $\mu$=4.4 \citep{smail99}. The SFR values for SMMJ2135-0210 were 
 corrected for lensing magnification using an amplification factor of $\mu$=32.5$\pm$4.5 derived by \citet{swinbank10}. The {\it Blue circles} show measurements of high-redshift far-infrared luminous galaxies 
 from \citep{swinbank04}. The gray dashed lines correspond to ratios of SFR(H$\alpha$) to SFR(FIR) of 0.01, 0.05, 0.1, 0.2, 0.3.\label{fig_starformation}}
\end{figure}

In Figure~\ref{fig_starformation}, we compare the SFR derived from the H$\alpha$ luminosity with the FIR SFRs for the galaxies in our sample, without correcting the H$\alpha$ measurements for 
extinction. We also include in this comparison a sample of 30 high-redshift far-infrared luminous galaxies from \citet{swinbank04}. The H$\alpha$ determined SFRs are significantly lower than the values 
determined from the IR luminosity, by factors of $\sim$3-5. This is most likely explained by both the presence of large amounts of dust obscuration in these systems and if there are fully-dust-obscured 
SF regions in the galaxy that do not contribute much to the Balmer emission lines. Unfortunately, the wavelength coverage of our spectroscopic observations does not extend to the $H\beta$ emission 
line, and thus the reddening in these galaxies cannot be estimated directly from the Balmer decrement, which would test this hypothesis. However, if we assume an average extinction value of 
A$_v$=2.9$\pm$0.5, as previously used by \citet{swinbank04} and measured by \citet{takata06}, we find that the extinction-corrected H$\alpha$ SFRs are in good agreement with those derived from 
the IR. The scatter seen in Figure~\ref{fig_starformation} could be explained by a combination of the fact that most of the star formation in these galaxies is not occurring in fully-dust-obscured regions, 
and due to the morphological diversity of sub-mm/ULIRGs-selected galaxies.

Estimates of the SFR from the H$\alpha$ flux in SMGs retain the substantial caveat that in the presence of an AGN, the blended nuclear emission may result in the broadening and brightening of 
the H$\alpha$ emission, potentially leading to overestimates of the SFR. Due to the large gas and dust reservoirs of SMGs, which could potentially provide ample fuel to trigger an AGN that could  
contaminate our estimates of SFR and $\rho$SFR. However, we note that only two sources in our sample show some evidence for the presence of an AGN, either via elevated [NII]/H$\alpha$ ratios or 
a broad line component. Previous IFU studies of SMGs have mainly targeted galaxies showing strong AGN signatures in their H$\alpha$ spectra \citep{menendez-delmestre13}, whereas the majority of 
our sample does not display clear H$\alpha$ AGN signatures. In SMGs, AGN were found to only contribute at low levels ($<$20\%) to the bolometric output \citep{alexander08}. However, even if an 
AGN is not dominating the bolometric output it may still affect the H$\alpha$ distribution and dynamics, being energetic enough to drive ionized gas over scales of a few kpc \citep{nesvadba08}.

\begin{deluxetable*}{lccccc}
\tablewidth{0pt} 
\tablecaption{Inferred physical properties derived from the VLT/SINFONI data.\label{tab_spec_der}}
\tablecolumns{6}
\tablehead{
\colhead{Name} &
\colhead{SFR(H$\alpha$)\tablenotemark{a}} &
\colhead{SFR(FIR)\tablenotemark{b}} &
\colhead{r$_{H\alpha}$} &
\colhead{$\Sigma$$_{SFR(H\alpha)}$} & 
\colhead{K$_{asym}$} \\
\colhead{} &
\colhead{[M$_{\odot}$yr$^{-1}$]}  &
\colhead{[M$_{\odot}$yr$^{-1}$]}  &
\colhead{[kpc]} &
\colhead{[M$_{\odot}$yr$^{-1}$kpc$^{-2}$]\tablenotemark{c}} &
\colhead{} 
}
\startdata
 J033246 &   26$\pm$4 & 692$_{-174}^{+174}$& 14.5 & 0.5 & 0.4926$\pm$0.0106\\
 J033249 & 132$\pm$17& 14084$_{-4520}^{+6750}$& 8.2 &7.9 & --\\
\hspace{12pt}   Region 1 & 58$\pm$6  & -- & 6.8 & 5.0 &  -- \\
\hspace{12pt}   Region 2 & 46$\pm$14 & -- & 5.8 & 5.5 &  -- \\
 RGJ0302+0010 & 171$\pm$22 & 1340$_{-244}^{+244}$ & -- & --  & 0.2380$\pm$0.0174\\
\hspace{12pt}   Region 1 & 73$\pm$25 & -- & 4.2 & 17.0  & --\\
\hspace{12pt}   Region 2 & 75$\pm$16 & --& 3.2 & 5.7 & -- \\
 L50879 &   60$\pm$8 & -- & 10 & 2.4 & --\\
 SMMJ04431+0210 (N4) &  77$\pm$20 &520& 11.5 & 2.3  & 4.3257$\pm$0.4890\\ 
\hspace{12pt} 	Region 1  &  30$\pm$4 & -- & 7.0 & 2.4 & --\\
\hspace{12pt} 	Region 2  &  20$\pm$3 & -- & 3.9 & 7.9  & --\\
 SMMJ2135-0120 & 82$\pm$4 & 398$_{-35}^{+35}$ & -- & -- &  0.2461$\pm$0.0018\\
\hspace{12pt} 	Region 1  & 33$\pm$7 & --&2.8 & 16.8 & --\\
\hspace{12pt} 	Region 2  &  54$\pm$8 & --&2.4 & 37.5 &  --\\
\hspace{12pt} 	Region 3  & 20$\pm$3 & --& 1.4 & 40.8 &  --
\enddata
\tablenotetext{a}{SFR(H$\alpha$) derived from the observed L$_{H\alpha}$ using the relation SFR(H$\alpha$) = 7.9 $\times$ 10$^{-42}$ L$_{H\alpha}$ [erg~s$^{-1}$] \citep{kennicutt98}. SFR(H$\alpha$) values 
are not corrected for extinction.}
\tablenotetext{b}{Far-IR luminosities are converted to SFR(FIR) using the relation SFR(FIR) = 4.5 $\times$ 10$^{-44}$ L$_{FIR}$ [erg~s$^{-1}$] \citep{kennicutt98}. The  SFR(FIR) of SMMJ2135-0120 and 
SMMJ04431+0210 (N4) are corrected for lensing magnification, assuming magnification factors of $\mu$=32.5$\pm$4.5 \citep{swinbank10} and $\mu$=4.4 \citep{smail99}, respectively.}
\tablenotetext{c}{Surface densities within r$_{H\alpha}$, determined, for instance, as: $\Sigma_{SFR(H\alpha)} = \frac{SFR(H\alpha)}{0.5 \times r_{H\alpha^{2}}}$.}
\end{deluxetable*}

\subsection{Schmidt-Kennicutt Relation}

We now use the H$\alpha$-derived sizes for these SMGs to compare the SFR surface densities determined from H$\alpha$ and FIR luminosities with the molecular gas surface density. Two of 
our sources have previously been observed in CO. Due to the uncertainties in the conversion between L'$_{CO(1-0)}$ and the H$_2$ mass \citep[e.g.,][]{bolatto13}, we consider 
both $\alpha_{CO}$ values of 4.6 and 0.8  K~km~s$^{-1}$~pc$^{-2}$~$^{-1}$, corresponding to spiral galaxies \citep{solomonybarrett91} and to ULIRGs \& star-forming galaxies at high 
redshift \citep{solomon05}, respectively.

For SMMJ04431+0210 (N4) we considered a line luminosity L'$_{CO(1-0)}$ 1.0$\pm$0.2 $\times$10$^{10}$ K~km~s$^{-1}$pc$^{-2}$ (corrected for the lensing magnification), and CO diameter of 
1.8 kpc, as reported by \citet{neri03}. With these measurements, we compute gas surface densities for $\alpha_{CO}$=0.8 and $\alpha_{CO}$=4.6 of 
3.1$\times$10$^{3}$ M$_{\odot}$pc$^{-2}$ and 
18 $\times$10$^{3}$ M$_{\odot}$pc$^{-2}$ respectively.  

For the galaxy SMMJ2135-0120 we use the line luminosities L'$_{CO(1-0)}$ reported by \citet{swinbank11} based on EVLA observations, corrected for lensing magnification. The reported luminosities 
are (3.9$\pm$ 0.4)$\times$10$^{8}$ K~km~s$^{-1}$pc$^{-2}$ for region Y1 and (7.1$\pm$ 0.8)$\times$10$^{8}$K~km~s$^{-1}$pc$^{-2}$  for region Y2. 
We use  the H$\alpha$ size measurements of 1.4~kpc for Region Y1 and 1.2~kpc for Region Y2. Then, we derive gas surface densities, assuming $\alpha_{CO}$=0.8 and $\alpha_{CO}$=4.6, of 6.3$\times$10$^{2}$ M$_{\odot}$pc$^{-2}$ and 3.6$\times$10$^{3}$ M$_{\odot}$pc$^{-2}$ for region Y1 and 1.5 $\times$10$^{3}$ M$_{\odot}$pc$^{-2}$ and 9$\times$10$^{3}$ M$_{\odot}$pc$^{-2}$ for region Y2, respectively and corrected for 
lensing magnification. For the other four sources, J033246, J033249, RGJ0302+0010 and L50879, there are no observations of CO emission available so far.

\begin{figure}
\begin{center}
\includegraphics[width=0.5\textwidth]{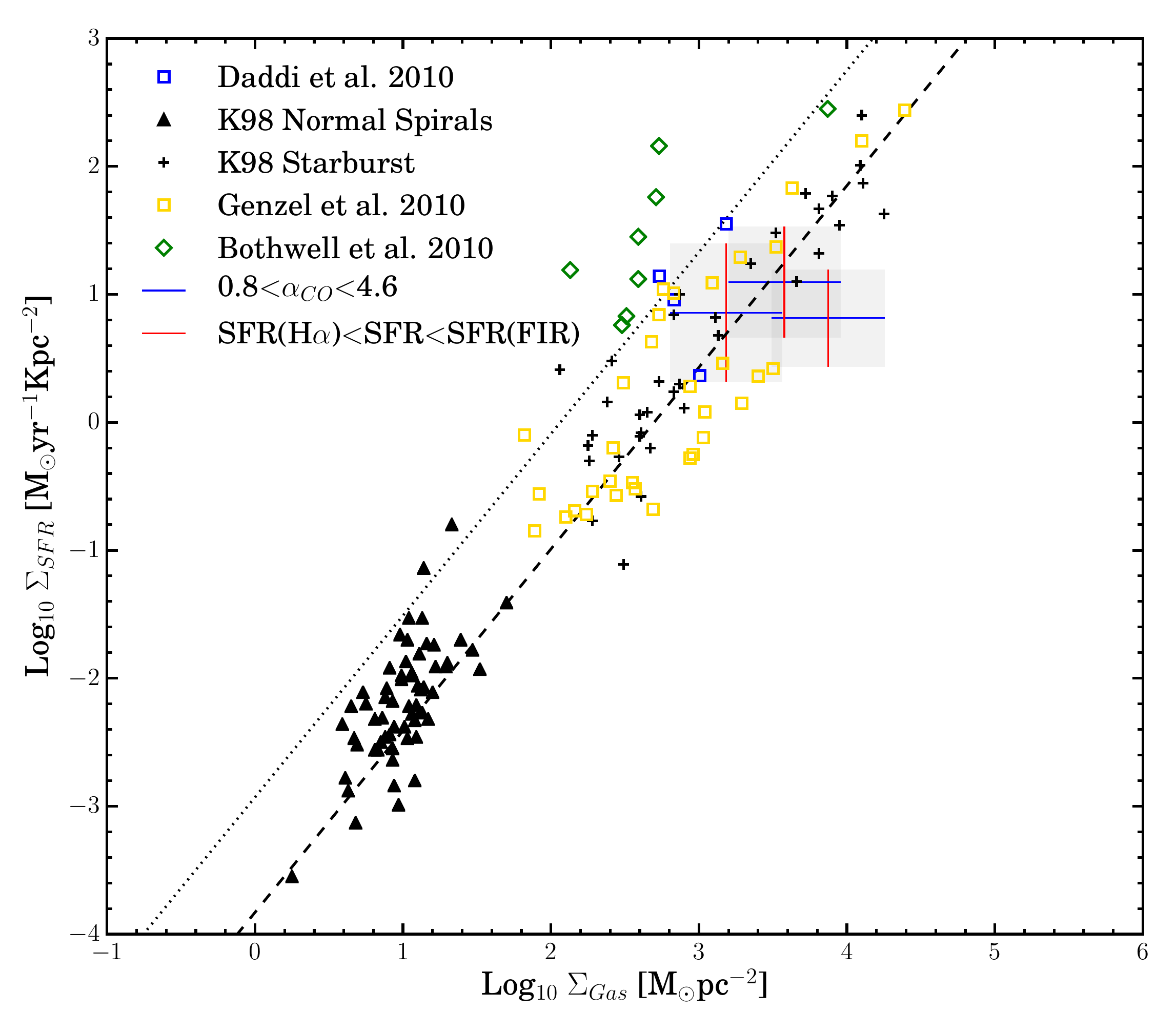}
\end{center}
\caption{SFR surface density versus gas surface density, both in logarithmic scale. {\it Blue lines} show the values for our sources using CO luminosity/mass conversion between 
$\alpha_{CO}$=0.8 and $\alpha_{CO}$=4.6. {\it Red lines} show measurements for the SMGs in our sample but using SFR between SFR(FIR) and SFR(H$\alpha$) instead. 
The {\it Blue squares} present the location of the BzK galaxies in the sample of \citet{daddi10b}, assuming a $\alpha$=3.2 value. {\it Yellow squares} show the SFRGs and SMGs in 
the \citet{genzel10} sample, using $\alpha_{CO}$=3.2 for SFGs and 1 for z$>$1 SMGs. {\it Green diamonds} show the SMG sample of \citet{bothwell10}, using $\alpha_{CO}$=0.8.
{\it Black triangles} show normal spiral galaxies, while the {\it Black pluses} present the starburst galaxies from the \citet{kennicutt97} sample, which were used to derive the relation 
shown by the {\it dashed black line}, a power law with a slope of 1.4. The {\it dotted black line} indicates the relation found by for \citet{daddi10}, a power law with a slope of 1.42 and
a higher normalization by 0.9 dex.}\label{fig_KS}
\end{figure}

Figure~\ref{fig_KS} shows the gas surface density versus the SFR surface density for the sources in our sample along with local and high-z sources from the literature.
The values from the literature were taken directly from each publication. As can be seen in Figure~\ref{fig_KS}, in all three sources for which we could measure $\Sigma$gas and $\Sigma$(SFR), 
they appear to be slightly below (but consistent) with the relation derived for local and high redshift star forming galaxies \citep[e.g.,][]{kennicutt97,genzel10}. This is in contrast
with previous results reported by \citet{daddi10b} and \citet{genzel10,genzel15}, who found a higher relation for both local and high redshift ULIRGs and SMG. While here we do not
attempt to correct the values previously reported for different assumed $\alpha_{CO}$ values, for our sources we assume a wide range of $\alpha_{CO}$, from 0.8 to 4.6, finding
that our sources still fall below the ULIRG and SMG relation and appear to be consistent with that for normal SFGs. While this is certainly interesting, it is hard to reach stronger conclusions
given the small sample studied here and the lack of high resolution CO data. Work to increase the number of sources in our sample and to obtain ALMA observations for them is
currently being carried out.

\section{Conclusions}
\label{sec:conclusions}

In this paper we used the VLT/SINFONI integral field spectrometer to study, at high resolution, the spatial distribution and kinematics of eight SMGs at 1.3$<$$z$$<$2.5 using the H$\alpha$ emission line
as a tracer. These observations allow us to study the gas dynamics and morphologies for six sources in our sample for which the H$\alpha$ line was detected at sufficient signal to noise ratio.

Overall, we find that these SMGs have highly irregular and/or clumpy morphologies. The H$\alpha$ emission has relatively large spatial extent, $\sim2-12$~kpc, with a mean value for our sample 
of $<$r$_{H\alpha}$$>$= 6.4~kpc, in contrast to local ULIRGs, which tend to be more compact. We find that in three cases the H$\alpha$ emission is significantly offset from the observed-frame 
near-IR continuum emission derived from the SINFONI and HST images.

We then analyzed the velocity and velocity dispersion fields in this sample, finding that these SMGs in general show large velocity gradients across each system. The majority of the SMGs in our sample are 
not consistent with being disk-like systems dominated by rotation and instead present irregular and turbulent or clumpy velocity and velocity dispersion fields, which could be explained by a recent major 
merger. We find that of the six SMGs with resolved spectroscopy, at least three appear to comprised of two or more dynamical sub-components. The average velocity offsets between these components 
is $\sim$180 km~s$^{-1}$ across a projected spatial scale of $\sim$8 kpc. The obvious merging/interacting nature of these systems suggests that they are analogous to the typically less luminous and 
slightly more compact ULIRGs in the local Universe. This is confirmed by a kinemetry analysis performed for the sources in our sample, which suggests that most of them do not show strong evidence 
for the presence of ordered motions.

The SMGs in our sample display high SFR surface densities ($\Sigma$$_{SFR}$), similar to those found in the most extreme local environments, such as circum-nuclear starbursts and IR-luminous 
galaxies but over larger spatial extents. Our IFU observations allow us to disentangle AGN and starburst-like components (from [N II]/H$\alpha$ flux ratios). Two of our targets (RJ0302-0010 and SMMJ04431+0210 (N4)) 
show possible signs of AGN activity surrounded by star-forming regions/clumps. We then further confirm that these extreme SMGs at high-$z$ appear to follow the same star-formation scaling relations 
as less luminous ``normal'' star forming galaxies. SMGs in our Group B sample are the only ones which present evidence for AGN activity. Therefore the data do not support the hypothesis that the 
AGN leads to hotter dust, at least for the sources in our Group A, which contain the warm-dust ULIRGs.

While still very expensive and limited in sample size, rest-frame optical IFU observations of high-$z$ extreme star-forming galaxies are starting to reveal the details of the violent star formation episodes in 
the early Universe. In the near future, such studies will be possible for much larger samples thanks to e.g., the KMOS multi-object IFU spectrograph at the VLT. In addition, ALMA observations will allow studies of the 
molecular gas contents, fuel for both star formation and for the central supermassive black holes, for these galaxies.

\acknowledgements
We thank the anonymous referee for very useful comments and suggestions. Support for this work was provided by the Center of Excellence in Astrophysics and Associated Technologies 
(PFB 06; ET, VO, FEB), by the FONDECYT regular grants 1120061, 1130528 and 1160999 (ET) and 1141218 (FEB), and by the ``EMBIGGEN'' CONICYT Anillo project ACT1101 
(ET, VO, FEB, GCP). We further acknowledge support from the FONDECYT Postdoctorado grant 3150361 (GCP) and the Ministry of Economy, Development, and Tourism Millennium 
Science Initiative through grant IC120009, awarded to The Millennium Institute of Astrophysics, MAS (FEB). IRS acknowledges support from STFC (ST/L007SX/1), the ERC advanced 
Grant Dustygal 3213324 and a Royal Society Wolfson Research Merit Award.


\end{document}